 \documentclass[preprint,3p,times]{elsarticle}
\usepackage{amsmath,amssymb,amsfonts,amsthm}
\usepackage{bbm}
 \biboptions{sort&compress}

\usepackage{float}
\usepackage[export]{adjustbox}
\usepackage{placeins}
\usepackage{booktabs}

\usepackage{amsmath,bm}
\usepackage{pdflscape}
\usepackage{url}
\usepackage{textgreek}
\usepackage[final]{microtype}
\usepackage{multirow}
\usepackage{caption}
\usepackage{subcaption}
\usepackage{wrapfig}

\usepackage[usenames]{color}

\usepackage{adjustbox}

\newcommand{\beginsupplement}{%
        \setcounter{table}{0}
        \renewcommand{\thetable}{S\arabic{table}}%
        \setcounter{figure}{0}
        \renewcommand{\thefigure}{S\arabic{figure}}%
        \setcounter{section}{0}
        \renewcommand{\thesection}{S\arabic{section}}
     }

\newcommand{\mb}[1]{\ensuremath{\mathbf{#1}}}

\journal{arXiv}

\begin{document}

\begin{frontmatter}

% \title{Optimizing roller-based spreading of fine, cohesive metal powders\\via DEM simulations}

%alterantive title #1
% \title{Novel Simulation-Inspired Roller Spreading Strategies for Fine and Highly Cohesive Metal Powders}

% \title{Coupled DEM-FEM simulation of improved methods for roller-based spreading of cohesive powders in additive manufacturing}

\title{Exploration of improved, roller-based spreading strategies for cohesive powders in additive manufacturing via coupled DEM-FEM simulations}

%alterantive title #2
%Uniform spreading of very fine, cohesive metal powders by novel simulation-guided roller kinematics and materials

\author[MIT]{Reimar~Weissbach\corref{cor1}}
\ead{reimarw@mit.edu}
\author[TUM]{Patrick~M.~Praegla}
\author[TUM]{Wolfgang~A.~Wall}
\author[MIT]{A.~John~Hart\corref{cor1}}
\ead{ajhart@mit.edu}
\author[TUM]{Christoph~Meier\corref{cor1}}
\ead{christoph.anton.meier@tum.de}

\address[MIT]{Department of Mechanical Engineering, Massachusetts Institute of Technology, 77 Massachusetts Avenue, Cambridge, 02139, MA, USA}
\address[TUM]{Institute for Computational Mechanics, Technical University of Munich, Boltzmannstra{\ss}e 15, Garching b. M{\"u}nchen, Germany}

\cortext[cor1]{Corresponding authors}

\begin{abstract}

Spreading of fine (D50 $\leq20~\mu m$) powders into thin layers, such as required by layer-wise additive manufacturing (AM) processes, typically requires a mechanism such as a roller that provides adequate shear and compression to overcome the cohesive forces between particles. Roller-based mechanisms that combine translation with rotation opposite to the direction of travel, i.e., counter rotation, are most commonly used. Roller-based spreading requires careful optimization and can result in low density and/or inconsistent layers depending on the specific characteristics of the powder feedstock. Here, we explore improved, roller-based spreading strategies for highly cohesive powders using an integrated discrete element-finite element (DEM-FEM) framework, which allows us to identify new spreading kinematics involving rotational oscillation of the roller and/or modifying the roller surface adhesion and compliance. Powder characteristics are emulated using a self-similarity approach based on experimental calibration for a Ti-6Al-4V 0-20~$\mu m$ powder used commonly in metal AM. We find that optimal roller-based spreading, quantified by maximum packing density and uniformity of layer thickness, relies on a combination of surface friction of the roller and roller kinematics, e.g., counter-rotation or angular oscillation, that impart sufficient kinetic energy to break cohesive bonds between powder particles. However, excess rotation (or shear stress) can impart excessive kinetic energy to the powder causing ejection of particles and a non-uniform layer, suggesting the existence of a process window of optimal spreading parameters. Interestingly, the identified optimal parameters for both investigated roller kinematics, i.e., angular velocity for counter-rotation as well as angular frequency and amplitude for rotational oscillation, result in a very similar range of roller surface velocities, suggesting this quantity as the critical kinematic parameter.
When these conditions are chosen appropriately, layers with packing fractions beyond 50~\% are predicted for layer thicknesses as small as $\sim$2 times D90 of the exemplary cohesive powder, and the layer quality is robust with respect to substrate adhesion over a 10-fold range. The latter is an important consideration given the spatially varying substrate conditions in AM due to the combination of fused/bound and bare powder regions. As compared to counter-rotation, the proposed rotational oscillation kinematics are particularly attractive because they can overcome practical issues with mechanical runout of roller mechanisms (which limit their precision). In particular, their application to rubber-coated rollers, which promise to reduce the risk of tool damage and particle streaking, is recommended for future experimental investigation.

\end{abstract}


\end{frontmatter}

\section{Introduction}

Powder bed-based metal additive manufacturing (AM) technologies are enabling complex, high-performance components to be realized for aerospace propulsion systems, medical implants, and industrial equipment, among many other applications~\cite{arabnejad2017fully,3DPrint2022}. Powder bed AM processes produce parts layer-by-layer by binding or fusing consecutive cross-sections. As such, the powder deposition process marks the first step in the periodically repeated sequence to build a part, and the powder layer quality has profound impact on the final part and overall process on many levels. Understanding how powder characteristics influence process outcomes and economics is therefore critical to continued industrialization of powder bed AM, and to accelerated development and application of new powder materials.

The two most prevalent powder bed-based metal AM processes are laser powder bed fusion (LPBF) and binder jetting (BJ). For both processes, the typical setup comprises a powder reservoir (typically a piston-actuated reservoir, or a moving hopper), a powder bed, and one or more spreading tools. In a typical LPBF machine, a piston raises the powder reservoir platform to provide a defined volume of powder as the first step. At the same time, the build platform is lowered by the nominal layer thickness to receive fresh powder. Next, the spreading tool distributes the powder from the powder reservoir onto the build platform, thus forming a new powder layer on top of the previous layer or substrate. After the powder spreading process is finished, a laser selectively fuses particles within the powder bed~\cite{meier2017thermophysical, MIAO2022103029}. 
For BJ, the setup commonly comprises a hopper to dispense the powder instead of a piston-actuated reservoir. After the hopper distributes an initial supply of powder over the build platform or in a continuous manner ahead of the spreading tool, the spreading tool creates a uniform and consistent layer on the build platform. Subsequently, the liquid binder is deposited using inkjet printing, bridging nearby particles in the desired cross-sections to create the 'green part' which is then removed from the printer and sintered to create the final part~\cite{li2020metal}.

In both processes, the powder layer influences (i) part quality as defects in the powder layer can propagate into the final part~\cite{MIAO2022103029}, (ii) production rate by means of the layer thickness, along with the spreading speed~\cite{li2020metal}, and (iii) the achievable level of detail of geometric features as limited by the particle size and layer thickness. In what follows, we briefly summarize the relevant background and state of the art for spreading of fine metal powders in AM, including computational modeling thereof.

\subsection{Powders used in LPBF and BJ} 

Powder bed-based metal AM typically uses a spherical powder produced by gas or plasma atomization as feedstock~\cite{Mukherjee2016, Gu2012}. Common materials include steel, nickel-based, titanium, copper, and aluminum alloys, and many ceramics~\cite{qian2015metal, li2020metal, avrampos2022review}, and the spherical particle shape is beneficial for uniform, consistent spreading~\cite{Boley2016MetalExperiment, Tan2017AnProcess}. Industrially produced powders usually can be modelled in good approximation with a lognormal particle size distribution (PSD). For LPBF, powders are typically chosen in a range within 10-60~$\mu m$ (D10 to D90, where DXX represents the XX-th percentile in the PSD) to achieve good spreadability along with a layer thickness that imparts high geometric resolution and good material quality as governed by the laser-material interaction during printing of each layer~\cite{Mindt2016, Vock2019, Brandt2017, Sutton2016}. For BJ, powders usually are chosen with a smaller mean diameter ($<$20~$\mu m$) to improve binder absorption by the powder bed and improve sintering characteristics, driven by the larger surface area per particle volume and the higher specific surface energy of smaller particles~\cite{li2020metal, bai2015effect, Ziaee2019}. Both processes benefit from using finer powders, but fine powders are difficult to dispense and spread into thin layers required for AM~\cite{Ziaee2019,Lores2019}.

Powder flow behavior depends on several factors, including size distribution, and especially on properties of individual particles, such as morphology or adhesive forces which is also influenced by humidity. When working with fine powders, adhesive forces dominate gravitational forces, leading to clumping and poor flowability~\cite{Walton2007, Walton2008, Herbold2015, Kloetzer2004ProcessCVD}, which generally results in non-uniformity and low packing fractions~\cite{Meier2019CriticalManufacturing} when using standard spreading mechanisms such as a rigid blade. Reuse of powder, which is required for reasons of process economics and sustainability, requires sieving/separating to remove spatter or satellite particle agglomerations generated by the melt pool dynamics~\cite{Cordova2019, avrampos2022review, fuchs2022versatile}. 

Given the significant challenges in working with fine powder, the majority of research focuses on spreading coarser, less cohesive powders~\cite{Wu2023} and existing literature consistently reports low powder bed quality and spreadability in the cases where more cohesive powders or smaller PSDs are investigated~\cite{MIAO2022103029,li2020metal}. To close this gap, we focus explicitly on very cohesive metal powders in this work.

\subsection{Spreading approaches in powder bed AM}

Machines for powder bed-based metal AM typically spread powder using a blade (most common for LPBF) and/or a roller (most common for BJ)~\cite{Haeri2017,Haeri2017_2,Snow2019}, to create layers with a typical thickness of $\sim$30-120$~\mu m$~\cite{Gibson2010, Vock2019}. The quality of the layer (e.g., uniformity, density, surface roughness), is primarily dependent on the feedstock properties and the geometry and kinematic motion of the spreading tool. 
%Layer thickness
Previous work has shown that the packing fraction of the powder bed increases with increasing layer thickness~\cite{yao2021dynamic,nan2020a, Meier2019CriticalManufacturing, zhang2020, Mindt2016}, and saturates at around 55-60\%.
More specifically, to avoid streaking of the layer, which is caused by the largest particles in the distribution, it has been shown that the nominal layer thickness should exceed $1.5\times D_{90}$, where $D_{90}$ is the 90\textsuperscript{th} percentile of particle diameters~\cite{Spierings2011}. Accordingly, the present study will employ a layer thickness of over $1.5\times D_{90}$. As a reference for typical packing fractions in high quality processes, an exemplary study reported 53\% average packing fraction for 28-48~$\mu m$ Ti-6Al-4V powder in a LPBF machine~\cite{Wischeropp2019}, which is sufficient for stable LPBF processing~\cite{Ziaee2019}.

%Traverse velocity
Rapid spreading is desirable to increase throughput, but the blade or roller velocity has been shown to have significant impact on the resulting layer quality. According to several studies, the packing fraction peaks for medium velocities and decreases significantly for high spreading velocities ($>$100$\frac{mm}{s}$)~\cite{Chen2017, Snow2019, yao2021dynamic, Haeri2017, chen2020packing, zhang2020}. Additionally, the surface uniformity of powder layers consistently decreases with increasing traverse velocity~\cite{Meier2019CriticalManufacturing, Parteli2016, Haeri2017, PENNY2022Blade, PENNY2022Roller}. For example, increasing the traverse velocity of a blade tool from 127$\frac{mm}{s}$ to 607$\frac{mm}{s}$ significantly decreased the packing fraction and surface uniformity of relatively coarse (PSD of 0-125~$\mu m$) polymer powders~\cite{Meyer2017}.
Potential sources for this behavior are bridging effects (i.e., particles build a stable force chain that 'bridges' the gap) and particle inertia both during feeding through the spreading gap as well as post-spreading inertia~\cite{chen2020packing, Meier2019CriticalManufacturing, Wang2021}.

%Tool
As mentioned above, common spreading implements used in powder bed AM are blades and rollers. For blades, which are typical in LPBF due to the larger and less cohesive powders used, round edges achieve higher packing fractions and higher uniformity than blades with a sharp edge~\cite{Meyer2017, Wang2021, Haeri2017}. Using a roller for spreading, with translation and counter-rotation, typically leads to a higher packing fraction than a blade under otherwise identical powder and spreading conditions, as shown by~\cite{Haeri2017, wang2020, shaheen2021influence, zhang2020, Budding2013, PENNY2022Roller}. This can be explained by a circulation of particles in the pile in front of the spreading tool, allowing a better rearrangement before being deposited~\cite{nan2020a}. Further factors such as roller diameter and roughness of the surface also influence the quality of the powder bed, as shown by Oropeza et al.~\cite{oropeza2022mechanized} who used X-ray microscopy to image the spatial distribution of packing density within layer spread with a mechanized apparatus. 

Forward-rotating roller kinematics have also been proposed as a means to densify the powder layer. However, there are considerable drawbacks associated with forward-rotation, primarily the significantly higher and strongly varying forces exerted on the powder bed. In LPBF, high local forces exerted by the spreading tool can damage thin and delicate printed features. In BJ, the structural integrity of the green part is low, leading to a risk of shear deformation~\cite{niino2009effect} or fracture of the weakly bound powder structure~\cite{li2021binder}. The powder bed can also be elastically deformed as a co-rotating roller passes, leading to a subsequent release of said elastic energy in the form of springback. This can lead to a thicker layer than intended and non-uniform surface profiles~\cite{ziaee2019fabrication}. A combination of counter-rotation and angular oscillation was first explored by Seluga~\cite{seluga2001} and showed promising results of improved packing fractions compared to pure counter-rotation when parameters are chosen correctly.
In \cite{nasato2021influence}, vertical oscillation is applied to DEM simulations of blade and roller spreading of a PA~12 powder. It was shown that for most scenarios, vertical oscillation of the spreading tool reduces powder layer quality.

The spreading tool can also become worn and damaged over repeated use, especially as consequence of non-uniform forces due to warpage of the part especially in LPBF, and interaction with agglomerates or solidified spatter. This damage, in turn, leads to tool vibrations resulting in layer defects~\cite{Vock2019, Hendriks2019}. Among others, this motivates the frequently used choice of compliant spreading tools, which accommodate high local forces with elastic deformation of the tool edge/surface. 
% Taken together, the dynamics of the spreading tool and uncertainties in powder properties (e.g., due to input variation, and implications of powder reuse) lead to a significant variance in the layer properties, namely the packing fraction. 

\subsection{Computational modeling of powder spreading}

Given the complex dynamics of powder spreading and the challenge of directly observing powder flow at the submillimeter length scales involved in spreading, computational modeling can lead to new insights that guide the development of practical methods and spreading mechanisms for improvement of AM processes. 
Pioneering works~\cite{Mindt2016, Herbold2015, Haeri2017, Gunasegaram2017} showed the suitability of discrete element method (DEM)~\cite{Luding2006,Luding2008} simulations to study powder spreading in LPBF. The authors showed in \cite{Meier2019CriticalManufacturing} the importance of considering cohesion for powder spreading, and a calibration strategy for the cohesive surface energy via angle of repose (AOR) experiments was presented in~\cite{Meier2019ModelingSimulations}. Referring to~\cite{Meier2019ModelingSimulations, Meier2019CriticalManufacturing}, several subsequent works adopted the approach to modeling and calibration of cohesion~\cite{Lee2020,HAN2019,wang2020,nan2020a,he2020,KOVALEV2020,zhang2020}. These previous contributions simulated spreading of polydisperse, spherical or non-spherical powders, considered visco-elastic contact, friction, rolling resistance, cohesion and fluidization~\cite{nan2020b} as particle interactions, and predicted layer properties such as packing fraction, surface uniformity, and particle segregation by investigating the influence of layer thickness, spreading velocity and spreading tool geometry~\cite{desai2019,shaheen2021influence}. It was shown that critical parameters are tool geometry, layer thickness, spreading kinematics, particle shape, and most importantly the level of cohesion, as described above. 

\subsection{Contribution of present work}

This paper presents an in-silico study of the complex multi-physics problem of roller-based spreading of highly cohesive powders, utilizing an integrated DEM-FEM framework. 
In particular, the computational studies explore roller kinematics including counter-rotation as well as angular oscillation applied to standard rigid rollers as well as coated rollers with compliant or non-adhesive surfaces. 
We find that optimal roller-based spreading, quantified by the packing density and uniformity of layer thickness, relies on a combination of surface friction between the roller and powder, along with roller kinematics that impart sufficient kinetic energy to break cohesive bonds between powder particles. However, excessive rotation can impart excessive kinetic energy to the powder causing ejection of particles and a non-uniform layer.
Interestingly, the identified optimal parameters for both investigated roller kinematics, i.e., angular velocity for counter-rotation as well as angular frequency and amplitude for rotational oscillation, result in a very similar range of roller surface velocities, suggesting this quantity as the critical kinematic parameter. When these conditions are chosen appropriately, layers with packing fractions beyond 50\% are predicted for layer thicknesses as small as $\sim$2 times D90 of the exemplary, highly cohesive powder, and the layer quality is robust with respect to substrate adhesion over a 10-fold range. The latter is an important consideration given the spatially varying substrate conditions in AM due to the combination of fused/bound and bare powder regions. As compared to counter-rotation, the proposed rotational oscillation kinematics are particularly attractive because they can overcome practical issues with mechanical runout of roller mechanisms (which limit their precision). In particular, their application to rubber-coated rollers, which promise to reduce the risk of tool damage and particle streaking, is recommended for future experimental investigation.

% When spreading conditions are chosen appropriately, packing fractions beyond 50~\% are predicted for layer thicknesses as small as $\sim$2 times D90 of the exemplary powder, and the layer quality is robust with respect to substrate adhesion over a 10-fold range. This is an important consideration given the spatially varying substrate conditions in AM due to the combination of fused/bound and bare powder regions. Moreover, parametric DEM-FEM simulations allow us to identify new spreading strategies involving angular oscillation of the roller rather than continuous rotation. We also study the effect of reduced roller adhesion and/or a compliant surface coating on the roller. Roller oscillation is particularly attractive because it can overcome practical issues with mechanical runout of roller mechanisms (which limit their precision), and is recommended for future experimental investigation.

The structure of this work is as follows. Section~\ref{sec:Methods} gives background on the central elements of the DEM-FEM model and describes the simulation setup, the quality metric as well as the employed model parameters. Subsequently, Section~\ref{sec:Results} reports the results of computational studies. Based on an initial comparison of blade vs. roller spreading for different levels of powder cohesion, one selected level of cohesiveness is investigated in the remainder of the study. For roller-based spreading with combined roller translation and counter-rotation, we assess the influence of traverse and rotational velocity as well as the roller friction coefficient. Based on these results, a process window for powder spreading is proposed. The results are then compared to an analysis of frequency and amplitude in spreading with angular oscillation, including spreading over a substrate with a reduced surface energy. The rotary and oscillatory cases are extended to scenarios with a roller with low surface adhesion, a roller coated with a flexible layer, and cases where the layer thickness is doubled, and the roller diameter is doubled. We conclude with an integrated perspective on our findings and possible future work.

\section{Methods}
\label{sec:Methods}

\subsection{Computational Model}

Spreading of fine (D50 $\leq20~\mu m$) metal powders is simulated via an integrated discrete element method-finite element method (DEM-FEM) framework that was first described in \cite{Meier2019ModelingSimulations}, implemented in the parallel multi-physics research code BACI~\cite{Baci}. The model has also been applied in~\cite{Meier2019CriticalManufacturing, Meier2021GAMM, Penny2021,PENNY2022Blade,PENNY2022Roller,Praegla2023,fuchs2021novel}, but the present work is the first in-depth study of roller spreading of fine powders. Each particle is modeled as a discrete spherical element. Structural components, such as the spreading tool or the substrate are modeled via FEM elements. Bulk powder dynamics are described with the resolution of individual particles following a Lagrangian approach. The model considers particle-to-particle and particle-to-wall (e.g., tool and substrate) interactions including normal contact, frictional contact, rolling resistance and cohesive forces.

Each discrete element (i.e., particle) has six degrees of freedom (i.e., three translational and three rotational DOFs) described by the position vector $\mb{r}_G$ and the rotation vector $\boldsymbol{\psi}$ and angular velocity $\boldsymbol{\omega}$ of the centerpoint of a particle. Based on these, the balance of momentum equations of a particle $i$ are~\cite{Meier2019ModelingSimulations}:
\begin{subequations}
\label{momentum}
\begin{align}
(m \, \ddot{\mb{r}}_G)^i = m^i \mb{g} + \sum_j (\mb{f}_{CN}^{ij}+\mb{f}_{CT}^{ij}+\mb{f}_{AN}^{ij}),\label{gusarov2007_HCE1}\\
(I_G \, \dot{\boldsymbol{\omega}})^i = \sum_j (\mb{m}_{R}^{ij}+\mb{r}_{CG}^{ij} \times \mb{f}_{CT}^{ij}),\label{gusarov2007_HCE2}
\end{align}
\end{subequations}
with the particle mass $m\!=\!4/3\pi r^3\rho$, the moment of inertia with respect to the particle centerpoint $I_G=0.4mr^2$, the particle radius $r$, the density $\rho$, and the gravitational acceleration $\mb{g}$. A bold symbol indicates a vector and a dot above a parameter indicates the derivative with respect to time of said parameter, e.g., $\dot{\boldsymbol{\omega}}^i$ is the angular acceleration vector of a particle $i$.
Each contact interaction between particles $i$ and $j$ is represented in Eq.~\eqref{momentum} through normal contact forces $\mb{f}_{CN}^{ij}$ (implemented with a spring-dashpot model), tangential contact forces due to Coulomb's friction $\mb{f}_{CT}^{ij}$ (with stick/slip frictional contact), adhesive forces $\mb{f}_{AN}^{ij}$, and torques $\mb{m}_{R}^{ij}$ due to rolling resistance. Furthermore, $\mb{r}_{CG}^{ij}:=\mb{r}_{C}^{ij}-\mb{r}_{G}^i$ is the vector pointing from the centerpoint of particle $i$ to the contact point with particle $j$. 

In~\cite{Meier2019ModelingSimulations}, a cohesion force law according to the Derjaguin-Muller-Toporov (DMT) model~\cite{Derjaguin1975}, was proposed and the interested reader is referred to that reference for more details on the formulation. The surface energy $\gamma$ is a critical parameter due to (i) high uncertainty about the magnitude and variance between particles, as it can vary by orders of magnitude due to variations in particle surface roughness/topology and chemistry; and (ii) high influence on the powder behavior such as flowability and spreadability. For example, for effective surface energy values on the order of $\gamma=1$e-4$~\frac{J}{m^2}$ as identified for representative powders~\cite{Meier2019CriticalManufacturing} and mean particle diameters in the range of $d=2r \approx 30~\mu m$, cohesive forces exceed gravitational forces by one order of magnitude.

The DEM-FEM coupling, which allows to model, e.g., deformable rollers, is described in detail in~\cite{praegla2024coupling}.

\subsection{Simulation setup}
\label{subsec:simulation setup}

\begin{figure}
\vspace{-12pt}
 \begin{center}
   \includegraphics[scale=1, keepaspectratio=true, width=\textwidth]{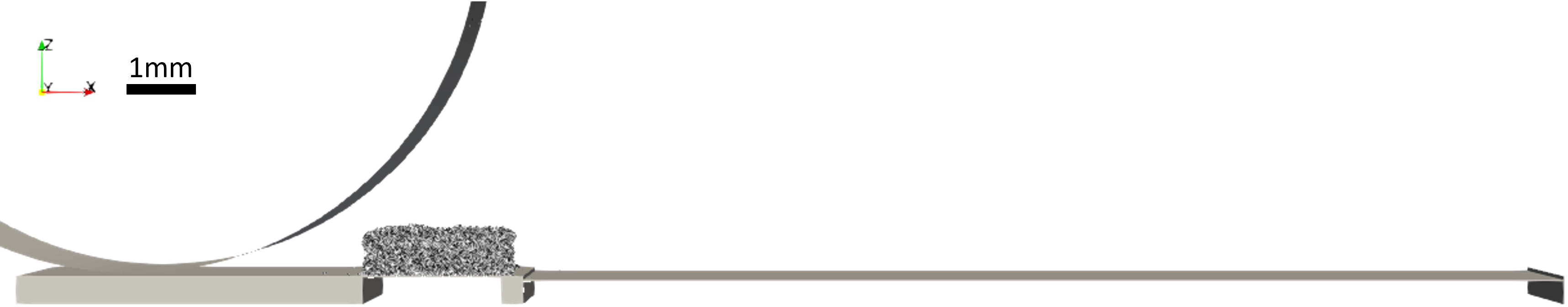}
 \end{center}
 \vspace{-16pt}
 \caption{Simulation setup with roller, fully extended powder reservoir piston with 42,000 particles and powder reservoir}
   \vspace{-12pt}
 \label{fig:simulation_setup}
\end{figure}

The following geometric setup is used for the simulation of powder spreading with a roller or a blade tool, with powder supplied by a reservoir as described below. The roller is modeled based on 125 linear circumferential segments of a circle with a diameter of 10~mm. The blade has thickness of 2~mm with a $90^{\circ}$ front edge. For roller simulations, 42,000 particles are used, whereas for the blade simulations, the number of 21,000 particles was used. For selected simulations that involve a larger roller with a diameter of 20~mm or a higher layer thickness, 105,000 particles are used. These numbers were chosen based on a sensitivity study and represent the threshold where a larger number did not significantly affect simulation results. The roller simulation requires a larger number of particles because the pile in front of and underneath the roller needs to be realistically scaled to the roller diameter. A larger roller thus requires more particles whereas a higher blade does not.
The dimensions of the powder reservoir and powder bed are similar to those used in previous work~\cite{Meier2019ModelingSimulations,Meier2019CriticalManufacturing, Meier2021GAMM, Penny2021,PENNY2022Blade,PENNY2022Roller}: the dimension perpendicular to the spreading direction is 1~mm with periodic boundary conditions, the powder bed has a length of 12~mm and quality metrics are evaluated over a 5~mm length, starting 3~mm after the beginning of the powder bed to avoid potential edge effects at the beginning or the end. The thickness of the layer is defined by the gap between the bottom edge of the spreading tool and the substrate, and is set to be $\sim$2 times the D90 particle diameter and just slightly thicker than the largest particle in the distribution (e.g., 100~$\mu m$ for a 15-45~$\mu m$ powder with a D90~=~43.4~$\mu m$ and cutoff diameter of 88~$\mu m$ or $\sim$40~$\mu m$ for a 0-20~$\mu m$ powder as described below). In the beginning of the simulation, particles are randomly allocated on a grid in the powder reservoir. The particles are released from the grid, dropping down due to the gravitational force to form a pile on the piston. Then, the powder reservoir piston moves the pile up and in front of the spreading tool. This situation is depicted in Figure~\ref{fig:simulation_setup}, just before the spreading tool starts to move across the powder bed.

Based on additional simulations at otherwise unchanged parameters, where out of the given log-normal distribution several powder samples were generated by means of different random seeds, we found that the resulting statistical variation of the results was significantly smaller as compared to the variation due to different spreading parameters (see supplemental Figure~\ref{fig:seed_vs_original}). Moreover, it has been verified that further increasing the contact stiffness as well as further decreasing the time step size (in the sense of a temporal convergence study) did not have a significant effect on the presented results.

\subsection{Quality metric: Spatially resolved packing fraction}

The primary layer quality metric analyzed in this work is the 2D packing fraction field $\Phi$ of the powder bed, quantified by its mean $\bar{\Phi}$ and standard deviation $std(\Phi)$. The computation and motivation of this metric is described in detail in \cite{Meier2019CriticalManufacturing} and will only be covered briefly here. This metric is assessed with a spatial resolution of 100x100~$\mu m$ in the length and width direction (denoted as bins), which is approximately the diameter of a typical laser spot in LPBF. For example, a low standard deviation of the packing fraction field indicates a high level of uniformity. The packing fraction is evaluated in two steps. First, each bin is further divided into smaller cubical 3D voxels with a side length of 2.5~$\mu m$ for a numerical integration of the particle volume. The voxel size is chosen as to minimize the error coming from the volume of particles cut by the segment walls. Finally, the packing fraction of the powder layer is evaluated as the ratio of the volume occupied by particles to the volume of the powder bed confined by the actual mean layer height. 

Note that for calculation of the packing fraction, the nominal layer height is commonly employed in the literature. We do not use the nominal layer height but the actual (calculated) mean layer height. This is important because the nominal layer height usually is considerably higher than the actual mean layer height of a spread layer. The mean layer height is calculated as the mean value of the layer heights of all 100~$\mu m$ bins~\cite{Meier2019CriticalManufacturing}.

\subsection{Powder model parameters}
\label{subsec:Powder model parameters}

To capture the powder size distribution in the simulations, the particle size distribution (PSD) was fitted to experimental laser diffraction measurements in our previous work~\cite{Penny2021}, and the cited values for the $10^{th}$-percentile (D10), median (D50) and $90^{th}$-percentile (D90) are fitted to a lognormal distribution in this work (D10~=~23.4~$\mu m$, D50~=~31.5~$\mu m$, D90~=~43.4~$\mu m$; cutoff diameter D$_{cutoff}$~=~88~$\mu m$). Most other powder model parameters are kept the same as referenced in~\cite{Penny2021,PENNY2022Blade,PENNY2022Roller}, and are listed in Table~\ref{tab:DEMparameters}.

Cohesive forces in the model are primarily determined by the surface energy $\gamma$ of the particles. A higher value of $\gamma$ defines a more cohesive powder. In \cite{Meier2019ModelingSimulations}, we showed that the critical dimensionless group that defines the cohesiveness of a powder is the ratio of gravitational and adhesive forces, which scales according to:
\begin{align}
    \frac{F_{\gamma}}{F_g} \sim \frac{\gamma}{\rho g r^2}.% \overset{!}{=} const.
\label{eq:Gravitation to Cohesion}
\end{align}
As shown in \cite{Meier2019ModelingSimulations}, the behavior of a given powder is similar as long as the ratio stays constant. For example, a powder with a lower density, such as an aluminum alloy, and a larger size distribution might behave similar to a higher density metal, such as steel, with a smaller size distribution.
Making use of this self-similarity condition, one can model the behavior of a powder based on the known behavior of others. The present model was calibrated on the 15-45~$\mu m$ Ti-6Al-4V powder (D10~=~23.4~$\mu m$, D50~=~31.5~$\mu m$, D90~=~43.4~$\mu m$; cutoff diameter D$_{cutoff}$~=~88~$\mu m$), with a surface energy value of $\gamma_0$~=~0.08~$\frac{mJ}{m^2}$~\cite{PENNY2022Roller,PENNY2022Blade}. 
In this work, we aim to model a very fine powder that has the same surface energy as the previously calibrated 15-45~$\mu m$ Ti-6Al-4V powder.
An exemplary commercial 0-20~$\mu m$ Ti-6Al-4V powder with D10~=~7~$\mu m$, D50~=~11~$\mu m$ and D90~=~18~$\mu m$ matches this requirement and can be modeled by retaining the size distribution of the 15-45~$\mu m$ powder, while using a modified surface energy in the model that is calculated with the transformation rule
\begin{align}
    \gamma_{fine} = \left(\frac{D50_{coarse}}{D50_{fine}}\right)^2 \times \gamma_{coarse}.
\label{eq:powder cohesion transformation}
\end{align}
Using this transformation rule, the modified surface energy takes on a value of $\gamma = 8~\gamma_0$ or 0.64~$\frac{mJ}{m^2}$, i.e., the surface energy is increased by a factor of 8 as compared to its consistent physical value.
The same surface energy used to describe particle-particle interactions is also applied to particle-boundary interactions. 
Using the radius of the median particle from the powder described above (D50~=~31.5~$\mu m$), as well as $\gamma$~=~0.64~$\frac{mJ}{m^2}$, $\rho$~=~4430kg/m$^3$, and $g$~=~9.8~m/s$^2$ and applying it to Equation~\ref{eq:Gravitation to Cohesion}, the dimensionless cohesiveness of the powder is approximately 60, meaning that cohesive forces dominate gravitational forces by a factor of 60 -- a highly cohesive powder.

\begin{table}
\centering
\renewcommand{\arraystretch}{0.7}
\begin{tabular}{ l l l } 
 \hline
 Parameter & Value & Unit \\ 
 \hline
 Density       &4430      &kg/m$^3$     \\ 
 Poisson's ratio     &0.342         &- \\
 Penalty parameter     &0.34         &N/m \\
 Coefficient of friction     &0.0-1.0         &- \\
 Coefficient of rolling friction     &0.07         &- \\
 Coefficient of restitution     &0.4         &- \\
 Surface energy     &0.02-2.56         &mJ/m$^2$ \\
 Hamaker constant     &$40\cdot10^{-20}$         &J \\
 Adhesion cut-off radius & $0.01$ & - \\
  \hline
 \end{tabular}
\caption{Default DEM model parameters}
\label{tab:DEMparameters}
\end{table}

Further, even though cohesive metal powders have a significantly different flow behavior than coarse granular media such as dry sand, flow behavior of metal powders can be generalized more easily due to their relatively similar, high stiffness~\cite{Mandal2020}. Therefore, the findings of this study can be applied to other materials of significance to LPBF and BJ, as well as other processes requiring spreading of metal powders.

\section{Results}
\label{sec:Results}

In what follows, we present parametric simulation studies focused on roller-based spreading of cohesive powders. We first compare blade versus roller spreading for powders with different levels of cohesiveness, then explore angular rotation, angular oscillation, as well as investigate the effect of reduced substrate adhesion and different modifications to the roller and simulation setup.

\subsection{Comparing roller and blade tools for spreading cohesive powders}

\begin{wrapfigure}{r}{0.5\textwidth}
% \vspace{-22pt}
 \begin{center}
   \includegraphics[width=0.5\textwidth]{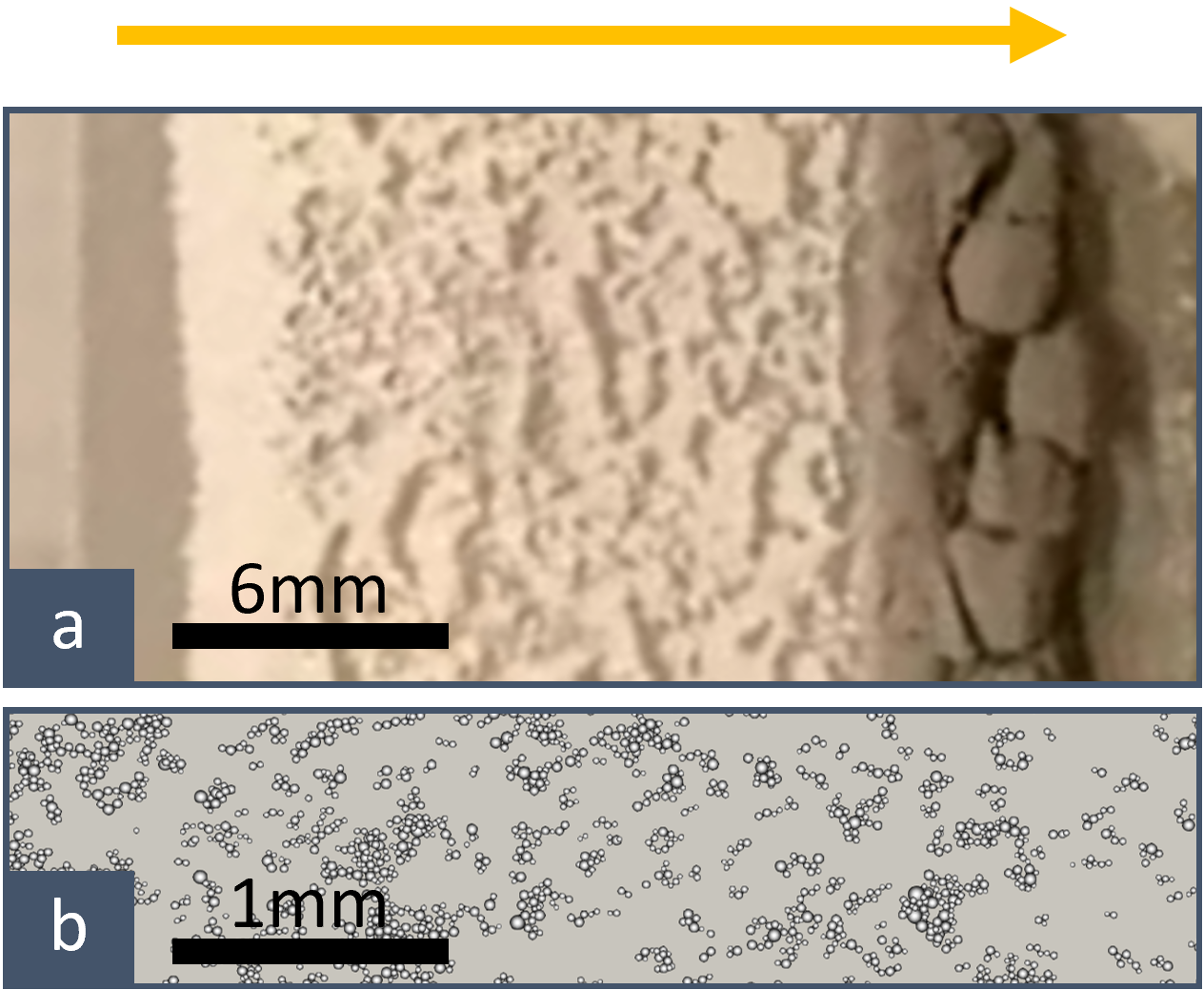}
 \end{center}
 \vspace{-16pt}
 \caption{Top view of Ti-6Al-4V 0-20~$\mu m$ powder spread with a blade a) experimentally and b) computationally}
    % \vspace{-12pt}
 \label{fig:sparse layer}
\end{wrapfigure}

To compare the blade and roller-based spreading via parametric simulations, the surface energy $\gamma$ was varied, spaced logarithmically between $0.25~\gamma_0 - 16~\gamma_0$ for blade spreading and $0.25~\gamma_0 - 32~\gamma_0$ for roller spreading. For spreading with traverse motion of a blade or (non rotating) roller, simulations predict deterioration of spreadability, as measured by a decrease in packing density, with increasing surface energy above $\frac{\gamma}{\gamma_0} = 1$. At the surface energy representing Ti-6Al-4V 0-20~$\mu m$ powder, the powder is unspreadable, with a layer packing fraction of approximately 10\%, with large void regions on the substrate after spreading. An experimental powder spreading result for the 0-20~$\mu m$ size distribution in a well of 100~$\mu m$ depth is depicted in Figure~\ref{fig:sparse layer}. Clearly, both the experiment and simulation show very sparse layers, only covered occasionally by clusters of particles. For quantitative experimental results for blade and roller spreading using a Ti-6Al-4V 15-45~$\mu m$ powder, the reader is referred to \cite{PENNY2022Blade} and \cite{PENNY2022Roller}.

\begin{figure}
     \centering
     \begin{subfigure}[b]{0.49\textwidth}
         \centering
         \includegraphics[width=\textwidth]{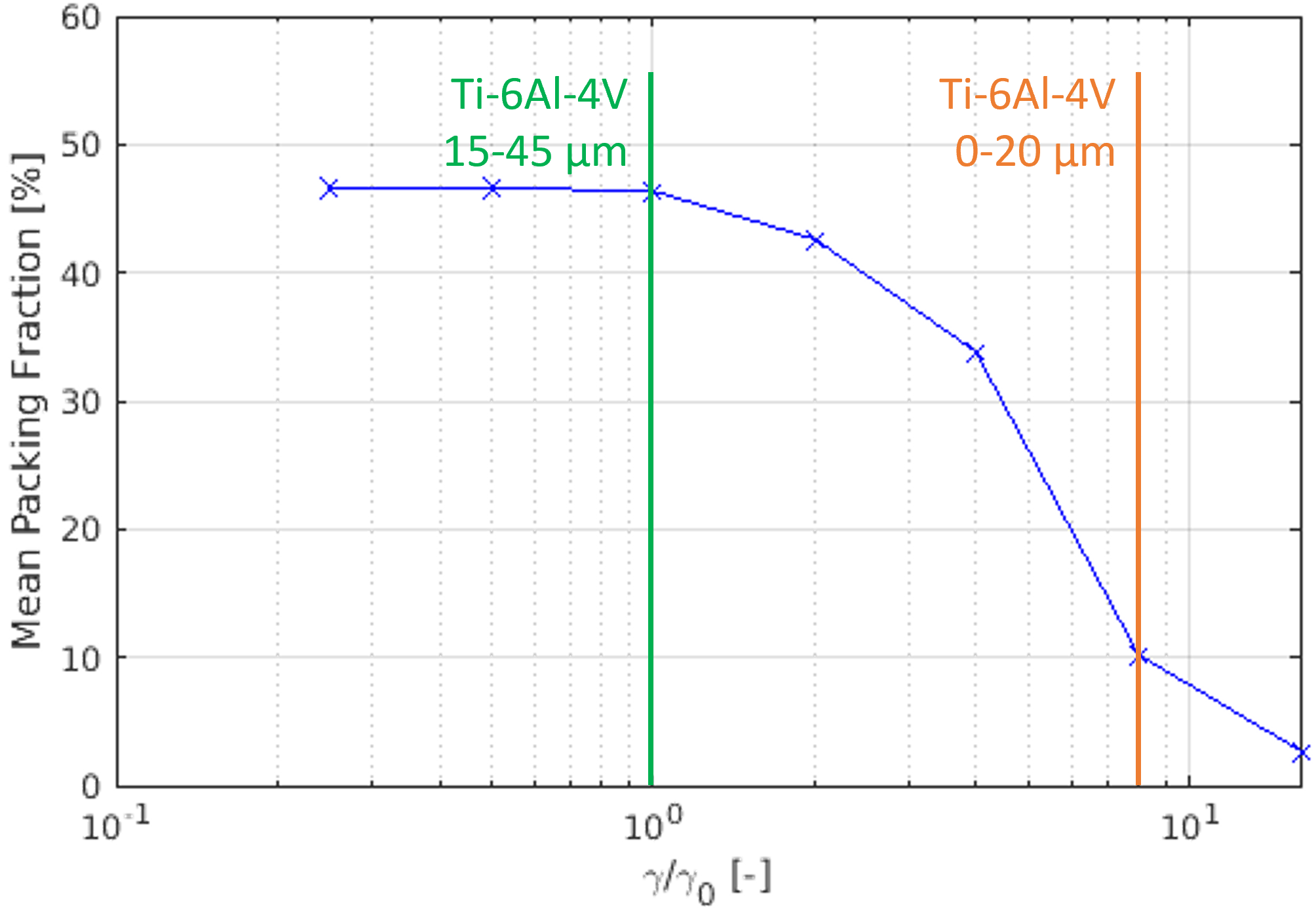}
         \caption{}
         % 90$^{\circ}$-blade spreading with traverse velocity of $50 \frac{mm}{s}$}
         \label{fig:Blade spreading}
     \end{subfigure}
     \hfill
     \begin{subfigure}[b]{0.49\textwidth}
         \centering
         \includegraphics[width=\textwidth]{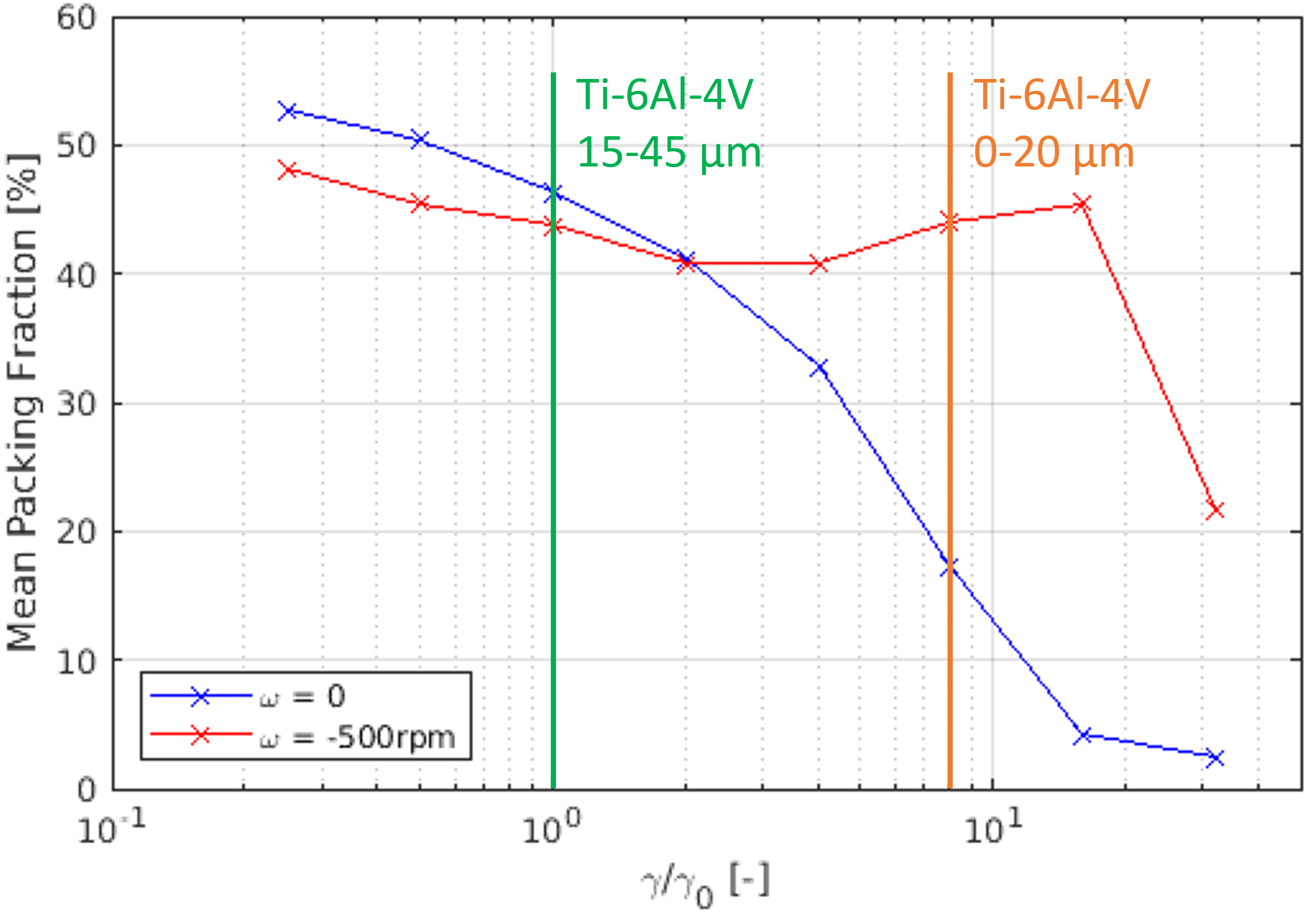}
         \caption{}
         % Roller spreading with traverse velocity of $50 \frac{mm}{s}$}
         \label{fig:Roller spreading vs cohesion}
     \end{subfigure}
     \caption{Simulated relationships between mean packing fraction of powder layer and normalized cohesion, for a traverse velocity of $50 \frac{mm}{s}$: (a) spreading with a 90$^{\circ}$ blade (based on data from~\cite{PENNY2022Blade}); (b) spreading with a roller, without rotation (based on data from~\cite{PENNY2022Blade}), and with counter-rotation at 500~rpm; Vertical lines indicate representative cohesion levels for typical LPBF and BJ powders with noted size distributions.}
     \label{fig:blade vs roller}
\end{figure}

For blade spreading, the expected deterioration of powder layer quality with increasing cohesiveness is observed in Figure~\ref{fig:Blade spreading}, quantified by mean packing fraction. The simulations were performed with a spreading velocity of $50\frac{mm}{s}$. Vertical lines on the plots indicate correspondence to levels of cohesion for representative powders used in LPBF and BJ, wherein blade and roller spreading approaches are typically used, respectively. The DEM parameter setting corresponding to the Ti-6Al-4V powder with a 15-45~$\mu m$ size distribution has previously shown very good agreement with experimental measurements of layer packing density by X-ray microscopy~\cite{Penny2021,PENNY2022Blade,PENNY2022Roller}. For the finer powders, the surface energy parameter was translated to the 0-20~$\mu m$ size distribution based on self-similarity considerations as described in Section~\ref{subsec:Powder model parameters}. 

When using a roller without rotation instead of the 90$^{\circ}$ blade, the general trends are the same; high packing fraction is achieved for non-cohesive powders, but with increasing cohesion the packing fraction continuously decreases, until powder is simply being swept across the build platform as one cohesive pile instead (see Figure~\ref{fig:Roller spreading vs cohesion}). The behavior significantly changes when the roller translates and counter-rotates, here at 500 rpm. While the packing fraction initially decreases with increasing surface energy, packing fraction increases again at surface energy above $4~\gamma_0$ and reaches a local maximum at approximately $16~\gamma_0$, before dropping suddenly.

The uptick in packing fraction for more cohesive powder is surprising, as even the highly cohesive powder corresponding to Ti-6Al-4V 0-20~$\mu m$, which was unspreadable in experiments using a standard blade or non-rotating roller implement~\cite{PENNY2022Blade,PENNY2022Roller}, shows good spreadability with the counter-rotating roller. Thus, the following sections will focus on the simulated Ti-6Al-4V 0-20~$\mu m$ powder corresponding to a dimensionless cohesiveness of 8.

\subsection{Powder-specific parameters for roller spreading}
\label{subsec:rotational spreading}

When spreading powder with a roller, AM practitioners usually can set the traverse (spreading) velocity $v$ and rotational velocity $\omega$. In addition, it is possible to select the roller material, surface texture, and coating to influence the powder-roller interaction. Collectively these may influence the roller-powder friction coefficient, $\mu$, which is modelled here.

\begin{figure}[h!]
     \centering
     \begin{subfigure}[b]{0.49\textwidth}
         \centering
         \includegraphics[trim = {8mm 0mm 10mm 2mm}, clip, scale=1, keepaspectratio=true, width=\textwidth]{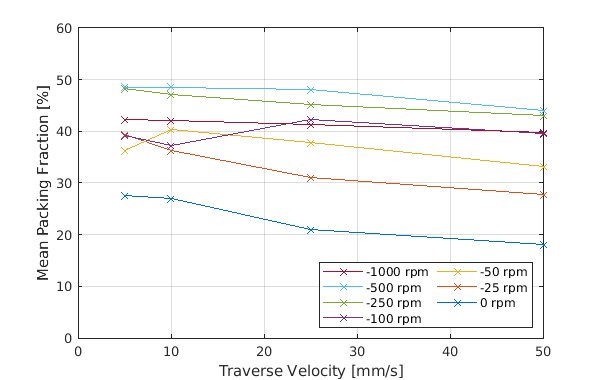}
         \caption{}
         \label{fig:Rotational study a}
     \end{subfigure}
     \hfill
     \begin{subfigure}[b]{0.49\textwidth}
         \centering
         \includegraphics[trim = {8mm 0mm 10mm 2mm}, clip, scale=1, keepaspectratio=true, width=\textwidth]{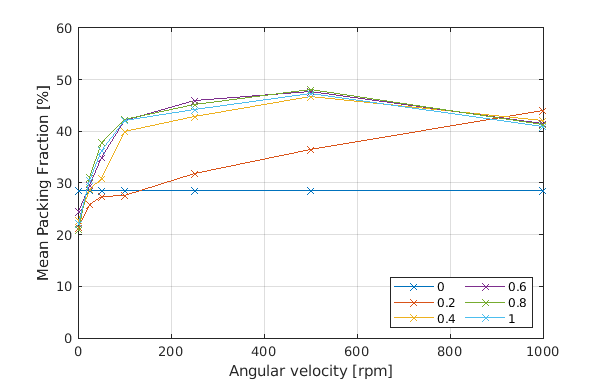}
         \caption{}
         \label{fig:Rotational study b}
     \end{subfigure}
     \vfill
     \begin{subfigure}[b]{0.49\textwidth}
         \centering
         \includegraphics[trim = {8mm 0mm 10mm 2mm}, clip, scale=1, keepaspectratio=true, width=\textwidth]{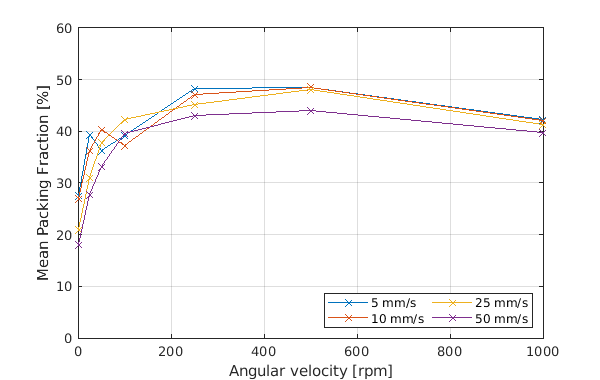}
         \caption{}
         \label{fig:Rotational study c}
     \end{subfigure}
     
     \caption{Simulated relationships between mean packing fraction and: (a) traverse velocity $v$ for different rotational velocities $\omega$ and roller-powder friction coefficient $\mu = 0.8$; (b) rotational velocity $\omega$ for different friction coefficients $\mu$ and traverse velocity $v=25\frac{mm}{s}$; (c) rotational velocity $\omega$ for different traverse velocities $v$ and roller-powder friction coefficient $\mu = 0.8$}
     \label{fig:rotational study}
\end{figure}

Therefore, the next parameter study was performed to elucidate the effects of traverse velocity, rotational velocity, and coefficient of friction on layer quality. Figure~\ref{fig:rotational study} shows an excerpt of the results, and the full data set is visualized for spreading velocities $v = \{5, 10, 25, 50\} \frac{mm}{s}$ in individual plots in Figure~\ref{fig:rotational study supp}. There, within each plot the mean packing fraction of a spread layer is shown for simulation results with $\mu = \{0, 0.2, 0.4, 0.6, 0.8, 1\}$ and $\omega = \{0, -25, -50, -100, -250, -500, -1000\}$~rpm. 

As mentioned earlier, spreading with forward roller rotation can lead to locally high forces perpendicular to the layer and substrate surface, which was confirmed by initial investigations. This can lead to both uneven layer surfaces as well as damage to e.g., the green part in binder jetting. For that reason, this study does not consider co-rotation as a viable spreading strategy. Nevertheless, we recognize that forward rotation can be used to increase packing density after spreading, and this is employed in some LPBF and BJ machines while it remains necessary to avoid excessive compaction forces.

There are several general trends observable in the data. First, as shown in Figure~\ref{fig:Rotational study a}, the packing fraction generally decreases with increasing roller traverse velocity. In some cases, the packing fraction first lightly increases (from $v=5\frac{mm}{s}$ to $v=10\frac{mm}{s}$), before it decreases again for higher velocities. This behavior of an optimal spreading velocity is in agreement with experimental results. Nan and Gu~\cite{nan2022experimental} argue that for very small spreading velocities (e.g., $v=1-5\frac{mm}{s}$), the shear strain rate is too small to sufficiently counteract the inter-particle cohesion to spread a dense powder layer. On the other hand, for very large traverse velocities, the inertia of particles can be a limiting factor as with increasing traverse velocity, the supply of powder from the pile to the gap might limit how much powder can be deposited by the tool. 
However, spreading velocity is the only parameter that directly relates to spreading time, and thus printing throughput. This means that a trade-off between throughput and quality has to be made, with the reduction in quality setting in around $25\frac{mm}{s}$ for the analyzed powder. For example, the highest packing fraction achieved for $v=10\frac{mm}{s}$ is 48.6\%, compared to 48.1\% for $v=25\frac{mm}{s}$ and 45.0\% for $v=50\frac{mm}{s}$.

The friction coefficient between the powder and roller also influences layer quality. At low values, the friction coefficient initially has a significant impact on the layer packing fraction, but beyond a friction coefficient of $\mu=0.4$, the impact is negligible, as shown in Figure~\ref{fig:Rotational study b}. $\mu \geq 0.4$ is common for normally machined steel roller surfaces~\cite{sekhar2018tunable}. In the theoretical case of $\mu=0.0$, no kinetic energy is transferred from the angular rotation to the powder pile, so the result is independent of the rotational velocity. With an increasing friction coefficient, the amount of shear induced by the angular rotation as well as the amount of kinetic energy transferred into the powder pile increases until saturation for a given kinematic parameter setting is reached. For $\mu=0.2$, the amount of kinetic energy transferred from the roller rotation to the powder is still relatively low and the data implies a linear correlation of packing fraction and angular velocity from 100~rpm to 1000~rpm. For values of $\mu \geq 0.4$, the friction coefficient does not seem to have a significant impact on the packing fraction. For all simulated values, the packing fraction reaches an optimum at $\omega=-500$~rpm. 

\begin{figure}[h!]
    \centering
    \includegraphics[width=\textwidth, scale=1, keepaspectratio=true]{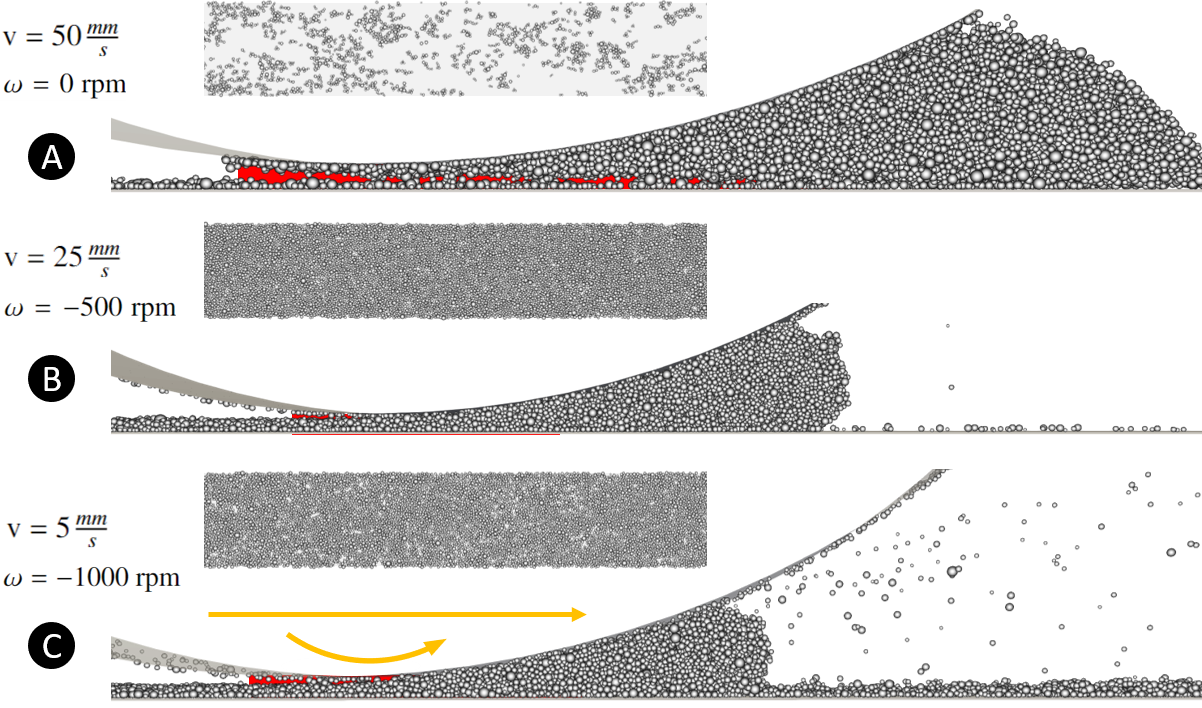}
    \caption{Visualization of counter-rotational spreading with $\mu = 0.4$, both in the side-view during spreading and top-view of the powder layer after spreading. Gap under the roller highlighted on red background. (A) v=$50 \frac{mm}{s}$, $\omega=0$~rpm; (B) v=$25 \frac{mm}{s}$, $\omega=-500$~rpm; (C) v=$5 \frac{mm}{s}$, $\omega=-1000$~rpm}
    \label{fig:process_window_images}
\end{figure}

The simulation data shows that rotational velocity is the most critical kinematic parameter, generally making the difference between forming a discontinuous powder layer and spreading a relatively dense and homogeneous layer with packing fractions up to 50\% (see Figures~\ref{fig:Rotational study c} and~\ref{fig:process_window_images}). Critically, the packing fraction saturates as the rotational velocity increases. Between $\omega=-250$~rpm and $\omega=-500$~rpm, only a small increase in packing fraction was observed and for the very high velocity of $\omega=-1000$~rpm the layer quality generally decreased again. Interestingly, for the higher angular velocity, even a low coefficient of friction is sufficient to achieve high packing fractions (see Figure~\ref{fig:Rotational study b}). The high rotational velocity paired with the low friction coefficient leads to a sufficient amount of shear in the powder pile to deposit the particles. For higher friction coefficients, the angular velocity of $\omega=-500$~rpm induces enough kinetic energy to break cohesive bonds while not excessively disturbing the powder heap (see Figure~\ref{fig:process_window_images}B). However, with higher angular velocity, the roller is creating dust and ejecting particles from the powder pile, leading to lower resulting packing fractions, as depicted exemplary in Figure~\ref{fig:process_window_images}C.

These results indicate that there exists a best-possible, powder-specific set of spreading parameters, for which even highly cohesive powder can be deposited successfully and with a high density and uniformity.

\begin{figure}[h!]
    \centering
    \includegraphics[width=.75\textwidth, scale=1, keepaspectratio=true]{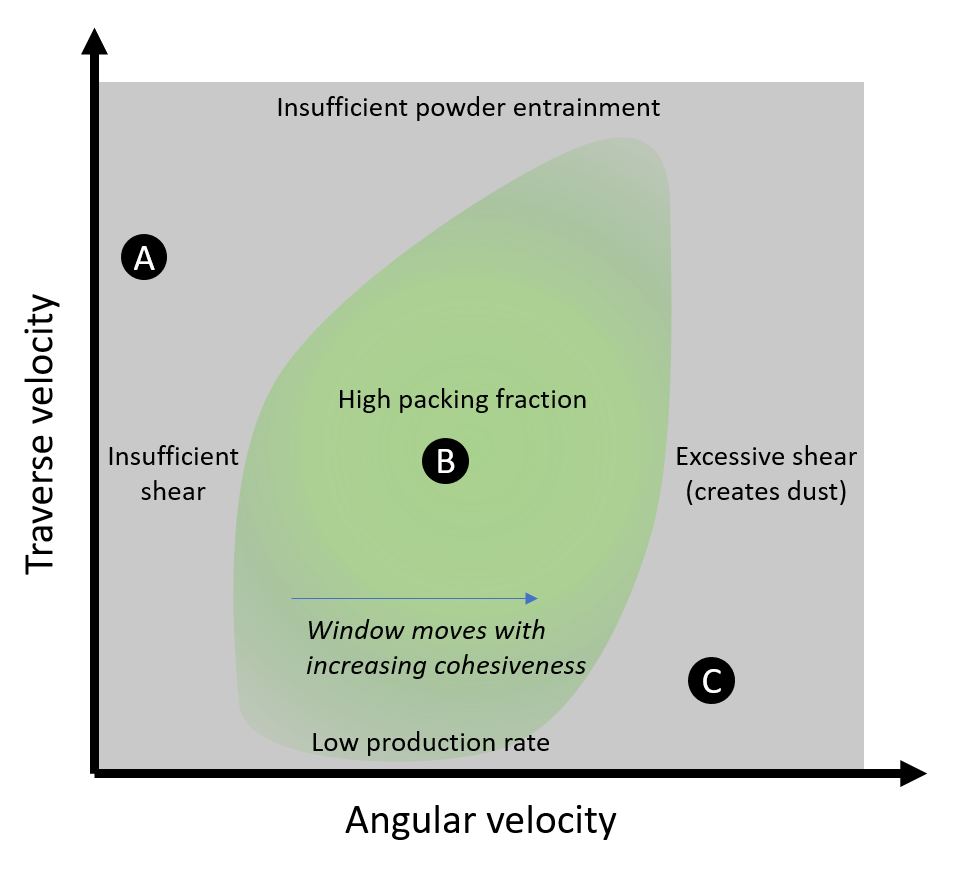}
    \caption{Schematic process window for spreading of cohesive powders; Exemplary simulation results from Figure~\ref{fig:process_window_images} are indicated with letters A-C}
    \label{fig:process_window}
\end{figure}

Generalizing these trends, we propose a spreading process window for roller spreading of cohesive metal powders. The window is made up by the two key parameters -- traverse velocity defining production rate (in addition to other steps of the AM process), and angular velocity inducing the majority of shear and kinetic energy into the powder pile, assuming a sufficiently high friction coefficient of $\mu \geq 0.4$. A high traverse velocity can constrain powder deposition due to the limiting factor of particle inertia in the pile and potential bridging effects as described earlier. For cohesive powders, the compressive force one the heap exacerbates the bridging effect as the pile forms a tightly packed cohesive cluster that is difficult to break. The supply of particles from the heap into the gap thus sets an upper limit on traverse velocity, as visible in Figure~\ref{fig:process_window_images}A, representing point A on the top left of the process window in Figure~\ref{fig:process_window}. In this simulation, the powder pile is simply pushed across the build platform and due to the lack of powder supply into the gap between the roller and the substrate, no shear is experienced by the powder in the gap (as shown by the 'void region' between particles adhering to the roller and particles adhering to the substrate). The resulting layer is very sparse. On the other side, very slow traverse velocities lead to low production rates which is often undesirable for economic reasons. Additionally, the shear component coming from the traverse velocity is reduced to a minimum as well.

The angular velocity adds the majority of the shear that the powder experiences in the gap (e.g., in the previously described simulation with v=$25 \frac{mm}{s}$, $\omega=-500$~rpm, the velocity component coming from the counter-rotation dominates the linear velocity component by an order of magnitude), but also induces kinetic energy into the heap in front of the roller, which can break cohesive bonds and 'free up' particles to flow into the gap. As such, there is a lower limit on the minimum angular velocity that is dependent on the level of cohesiveness of a given powder. Less cohesive powders generally will have a lower limit, with free-flowing powders not necessarily requiring angular rotation at all. The amount of kinetic energy induced into the powder heap also sets an upper limit on angular velocity. As shown above, at a certain point, the angular rotation transfers such an excess amount of kinetic energy, that the roller ejects powder particles from the heap, creating a 'dusty zone' in front of the roller. This is visible in Figure~\ref{fig:process_window_images}C, representing point C in the process window in Figure~\ref{fig:process_window}. For the simulation corresponding to point C (v=$5 \frac{mm}{s}$, $\omega=-1000$~rpm), the rotation ejects particles with such velocity, that the powder pile is not able to fully fill the gap between the roller and the substrate. This leads to low packing fractions and a less uniform powder layer. %With a significantly increased traverse velocity for the simulation corresponding to point D (v=$50 \frac{mm}{s}$, $\omega=-1000$~rpm), the powder pile is able to close the gap better because the inertia of the pile in combination with the agitation coming from the angular rotation actually helps supplying powder to the gap in this case.
Balancing all of these effects, a parameter combination from the center of the process window can create a consistent powder layer with a high packing fraction. This is exemplified with simulation B, with v=$25 \frac{mm}{s}$, $\omega=-500$~rpm.

While requiring further investigation, the results suggest that for an increasing level of cohesion in the powder, the process window will move further to the right and get more narrow. More cohesive powder requires a higher amount of shear and kinetic energy/agitation to break cohesive bonds and supply powder to the gap, while the sensitivity to optimal spreading parameters likely increases. For the theoretical case of non-cohesive powder, no angular velocity is required to spread powder with a high packing fraction, given a moderate traverse velocity (see e.g., \cite{Meier2019CriticalManufacturing}), but the powder is also more robust to higher angular velocities as can be seen in Figure~\ref{fig:blade vs roller}b and~\cite{PENNY2022Roller}. 
On the other hand, the minimum required angular velocity increases with an increasing level of cohesiveness in order to break the bonds between particles -- this is likely dictated by the pull-off force that keeps particles attached to each other. The upper limit where particles or particle clusters are ejected from the pile in front of the roller does not move as fast to the right since the energy required to eject particles is a function of their mass (overcoming gravity and inertia) as well as cohesiveness (pull-off force from adjacent particles). With an increasing level of cohesion, the relevance of inertia/gravity decreases.
Building on this analysis and by using powder rheology, future work could define a normalized process window that correlates process parameters to spreadability.

Interpreting the results, we also want to draw an analogy between powder spreading with a roller and the conventional manufacturing techniques grinding or milling. During high speed grinding for example, the contact zone where the tool interacts with the work piece is characterized by a viscous shear layer~\cite{leopold2000challenge}. This layer is comparable to the resulting shear zone when spreading a cohesive powder pile with a counter-rotating roller. For a given traverse velocity, a higher cutting velocity -- caused by a higher rotational velocity in peripheral milling and grinding processes -- is known to reduce the amount of material removed per rotation, leading to decreased cutting forces and an increase in surface quality~\cite{neugebauer2011velocity, kalpakjian2009manufacturing}, effects that are also observed for the spreading of very cohesive powders using a counter-rotating roller. This holds true until the effect of particle ejection is observed.

\subsection{Roller-based spreading using angular oscillation}
\label{subsec:oscillation spreading}

Although spreading with a counter-rotating roller can be highly effective, practical outcomes are constrained by the manufacturing precision of the spreading apparatus~\cite{PENNY2022Roller}. Roller runout -- defined as non-circularity in the position of any reference point on the surface of the roller, as it rotates -- can impart a wavy profile in the powder layer, depending on the relative rotation and translation kinematics (see Figure 3 in~\cite{PENNY2022Roller}). Runout is caused by native imprecision in the roller geometry and its rotation, which is governed by the tolerances of machining, and those of mechanical components such as rolling element bearings as well as the overall assembly~\cite{oropeza2022mechanized}. As an approximation, for intended powder layers no greater than 100~$\mu m$ thickness, roller runout of only 10~$\mu m$ can significantly impact the local layer thickness and therefore the effective packing density. 

To potentially address this consideration, we used the simulation framework to study roller-based spreading with small-amplitude angular oscillations. By using only a segment of the roller surface, the effect of manufacturing runout is mitigated, as local variations over a small sector of the roller circumference are typically much less than the total circular runout.

Seluga in 2001 proposed for the first time using angular oscillation to assist powder spreading~\cite{seluga2001}. In that work, a cohesive 0-15~$\mu m$ 17-4PH~SS powder was coated with a polymer to make it less cohesive and spreadable. A custom-build spreading apparatus was used, with traverse velocities between 3-12$\frac{mm}{s}$ and an angular velocity of -60~rpm (i.e., counter-rotating). This motion was superimposed by a 60~Hz angular oscillation with varying amplitude. The superposition of the two rotational movements led to two components of the relative velocity in the spreading gap. For small oscillation amplitudes (around 0-1~mm on the roller surface), no effect on the packing fraction was observed. For higher amplitudes that lead to the relative velocity due to oscillation being larger than the relative velocity due to steady traverse and rotational movement, the packing fraction was significantly improved (amplitudes around 1.5-2.5~mm on the surface). However, the surface of the powder layer was damaged with ridges due to a net-forward rotation of the roller during parts of the oscillation period. These ridges were found to be as deep as 30~$\mu m$ in a 50~$\mu m$ nominal powder layer, effectively rendering the layers unsuitable for LPBF or BJ processing.

Instead of a constant rotational velocity $\omega$, an oscillating rotational velocity that depends on the amplitude $A$, the frequency $f$ and the time $t$ is defined as:
\begin{align}
    \omega (t) &= A~2\pi f~\cos(2\pi f~t).
    \label{eq:oscillation_omega}
\end{align}
This definition of the angular velocity implies an angular displacement $\alpha$ of a specific point on the roller surface with the maximum amplitude $A$ which is defined as:
\begin{align}
    \alpha (t) &= A~\sin(2\pi f~t).
    \label{eq:oscillation_alpha}
\end{align}

This angular oscillation can be interpreted in two ways. First, the frequency of the oscillation can indicate the regime, e.g., above 20 kHz is commonly defined as ultrasonic vibration. Alternatively, the maximum (absolute) circumferential velocity can be related back to an equivalent rotational velocity if the oscillation's maximum speed is taken as the velocity of a constant rotation. That equivalent constant rotational velocity can be easily understood in context of e.g., the results in Section~\ref{subsec:rotational spreading}. 

In the simulations that follow, the amplitude and equivalent rotational velocity are being kept as independent variables and the frequency is chosen as dependent variable. The smallest amplitude is chosen to be approximately half of the D50 of the PSD, which corresponds to 0.08$^{\circ}$. The other examined amplitude values are 1$^{\circ}$ (i.e., 87~$\mu m$ amplitude on the roller surface, corresponding to the cutoff diameter of the PSD) and 10$^{\circ}$ (i.e., 873~$\mu m$ amplitude on the roller surface). For the equivalent rotational velocity, values ranging from 10~rpm to 1000~rpm are chosen, which cover approximately the same range as in Section~\ref{subsec:rotational spreading}, while the traverse velocity is constant for all simulations at $25\frac{mm}{s}$, which was identified as best compromise between spreading accuracy and throughput in Section~\ref{subsec:rotational spreading}. Further, a friction coefficient of $\mu = 0.8$ is chosen for all following simulations as it produced the highest packing fractions values. Figure~\ref{fig:rotational study} shows however, that the results for friction coefficients above $\mu = 0.4$ don't meaningfully differ. The full set of parameters is given in Table~\ref{table:oscillation parameters}.

\begin{table}
\centering
\caption{Frequencies for amplitude-'equivalent maximal rotational velocity' combinations of the parameter study}
\label{table:oscillation parameters}
\begin{tabular}{ |p{3.7cm}|r|r|r| } 
 \hline \raggedleft
  Equivalent max rotational velocity [rpm] &Frequency for: 0.08$^{\circ}$, [Hz] &for 1$^{\circ}$, [Hz]           &for 10$^{\circ}$, [Hz]\\ 
 \hline
 \raggedleft10    & 119   & 9.55& 0.955 \\ 
 \raggedleft50    & 595   & 47.7 & 4.77\\ 
 \raggedleft100   & 1190      & 95.5 & 9.55\\ 
 \raggedleft250   & 2975      & 239 & 23.9 \\
 \raggedleft500   & 5950      & 477 & 47.7 \\ 
 \raggedleft750   & 8925      & 716 & 71.6 \\ 
 \raggedleft1000  & 11900    & 955 & 95.5 \\
 \hline
\end{tabular}
\end{table}

\begin{figure}[h!]
     \centering
     \begin{subfigure}[b]{0.49\textwidth}
         \centering
         \includegraphics[trim = {8mm 0mm 10mm 2mm}, clip, scale=1, keepaspectratio=true, width=\textwidth]{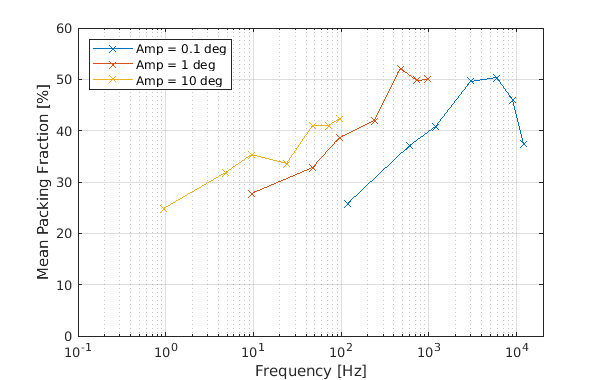}
         \caption{Results plotted by oscillation amplitude}
         \label{fig:Oscillation_amp}
     \end{subfigure}
     \hfill
     \begin{subfigure}[b]{0.49\textwidth}
         \centering
         \includegraphics[trim = {8mm 0mm 10mm 2mm}, clip, scale=1, keepaspectratio=true, width=\textwidth]{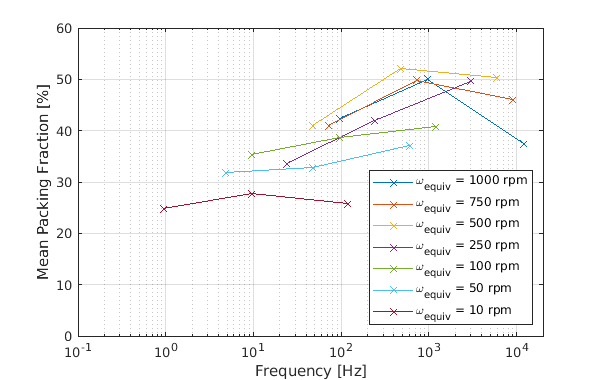}
         \caption{Results plotted by max circumferential and corresponding rotational velocity}
         \label{fig:Oscillation_vmax}
     \end{subfigure}
     \caption{Mean packing fraction of powder layers spread with angular oscillation for different frequencies and amplitudes}
     \label{fig:oscillation study}
\end{figure}

Figure~\ref{fig:oscillation study} shows the results of the parameter study from two different perspectives. The data points in Figure~\ref{fig:Oscillation_amp} and Figure~\ref{fig:Oscillation_vmax} are the same, but they are grouped differently. 
Looking at Figure~\ref{fig:Oscillation_amp}, it is evident that for each amplitude, an increase in frequency generally increases the packing fraction until a plateau is reached. Similarly to the pure rotational spreading in Section~\ref{subsec:rotational spreading}, the plateau is followed by a decline for very high frequencies. 
Further, the highest packing fractions are achieved with a medium amplitude around 1$^{\circ}$. This is visible both in Figure~\ref{fig:Oscillation_amp} as well as by the 'hat' shape of the curves in Figure~\ref{fig:Oscillation_vmax} where the middle data point of each curve represents the 1$^{\circ}$-amplitude data point.
From Figure~\ref{fig:Oscillation_amp} it can be concluded that employing a smaller amplitude requires a higher frequency to reach similar results. Shifting the perspective to Figure~\ref{fig:Oscillation_vmax} helps understand this behavior better as each amplitude-frequency combination is mapped to an equivalent maximal rotational velocity. With increasing equivalent maximal rotational velocity, the packing fraction increases, up to 500~rpm. Beyond the peak at 500~rpm, the packing fraction decreases again, similarly to the pure rotational simulations.
Overall, the highest packing fractions are achieved for values between 0.5-5 kHz and amplitudes of 1$^{\circ}$ or below, which corresponds to equivalent maximal rotational velocities around 500~rpm with the employed roller geometry.

\begin{figure}
% \vspace{-12pt}
 \begin{center}
   \includegraphics[width=\textwidth]{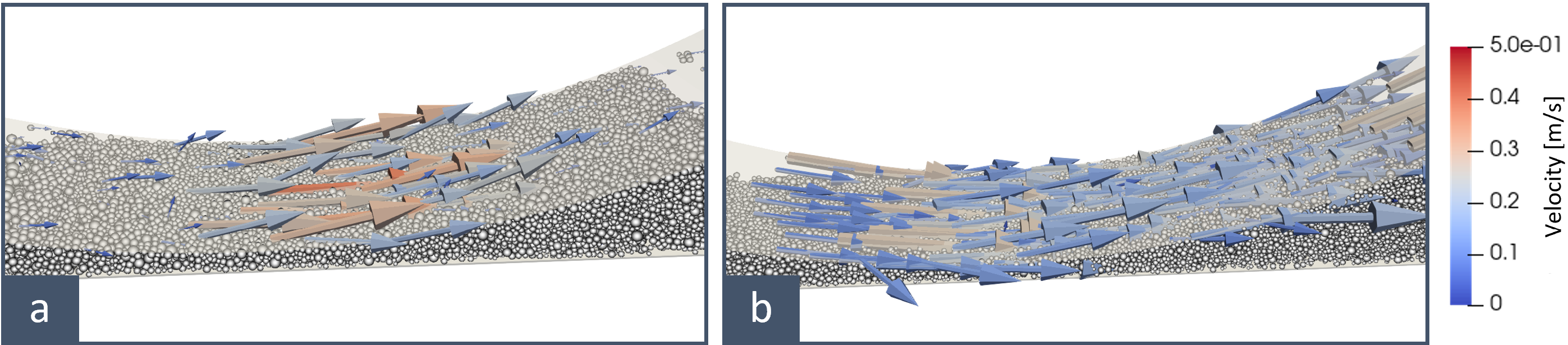}
 \end{center}
 \vspace{-16pt}
 \caption{Comparison of powder particle velocities for a) 1000~rpm equivalent with 0.08$^{\circ}$ amplitude and b) 500~rpm equivalent with 1$^{\circ}$ amplitude}
   % \vspace{-12pt}
 \label{fig:high oscillation comparison}
\end{figure}

Qualitative analysis of the particle movement revealed that for very high frequencies and small amplitudes, corresponding to an equivalent velocity above 500~rpm, the inertia of the powder is too high for the roller oscillations to affect the particles. Figure~\ref{fig:high oscillation comparison} shows the exemplary comparison of the highest frequency oscillation (1000~rpm equivalent, 0.08$^{\circ}\approx$~0.1$^{\circ}$ amplitude) and the best layer (500~rpm equivalent, 1$^{\circ}$ amplitude). The arrows indicate the velocity of each particle, with the color being on the same scale as shown in the legend. The length of the arrows for the 1000~rpm case is scaled by a factor of 0.5 compared to the arrows of the 500~rpm case in order to account for the difference in reference velocity of the roller (1000~rpm vs. 500~rpm). In the high frequency case, only particles that are vertically underneath the centerpoint of the roller are affected by the angular movement. This is a typical behavior for a non-rotating/non-oscillating roller where particles in the narrowest part of the gap experience high forces and particles outside that region are largely unaffected. For the best case in Figure~\ref{fig:high oscillation comparison}b, most particles contacting the roller experience a similar velocity (many similarly sized/colored arrows). This even velocity field indicates strong engagement of the roller with the powder pile. The oscillating nature then breaks cohesive bonds between particles and creates a relatively uniform surface. 

Comparing oscillating spreading to pure rotational spreading, several additional differences are visible. For rotational spreading, the roller surface is completely covered with powder as it adheres to the roller. This effect is also visible in experimental studies performed in our lab. For oscillating spreading, only a small part of the roller surface interacts with the powder and thus has particles adhering to it. Potentially related to that, rotational spreading shows significant ejection of particles from the spreading zone forward onto the powder bed. This creates powder dust as the roller moves, especially for high angular velocities. With the oscillating spreading, almost no such dust creation was visible in the simulations. These effects are in addition to the inherent benefits of oscillation such as no roller runout problem, and the fact that no 'full' roller is required, but only a roller segment.

\subsection{Influence of reduced substrate adhesion}
\label{subsec:reduced sadh}

During LPBF, interlayer powder spreading occurs over an area that is a combination of previously spread powder and melt tracks. The surface properties of powder and melt tracks are quite different, presenting different local boundary conditions for spreading of the next powder layer. To evaluate the influence of a change in surface properties of the underlying surface (i.e. substrate, powder bed or melt track), the substrate surface energy, as critical parameter for the spreadability, is reduced, which makes the spreading more challenging and allows us to assess the robustness of the different spreading strategies. While it is challenging to asses how much the actual surface properties (friction coefficient, contact surface, etc.) differ between melt track and powder bed, we simply analyze a very challenging case. In this section the substrate surface energy levels are reduced by a factor of 10, meaning that: $\gamma_{substrate, modified} = 0.1\gamma$. For this changed setting, the parameter studies from Section~\ref{subsec:rotational spreading} (for traverse velocity of v = $25 \frac{mm}{s}$) and Section~\ref{subsec:oscillation spreading} (also v = $25 \frac{mm}{s}$) are repeated. 

\begin{figure}[h!]
     \centering
     \begin{subfigure}[b]{0.49\textwidth}
         \centering
         \includegraphics[trim = {8mm 0mm 10mm 2mm}, clip, scale=1, keepaspectratio=true, width=\textwidth]{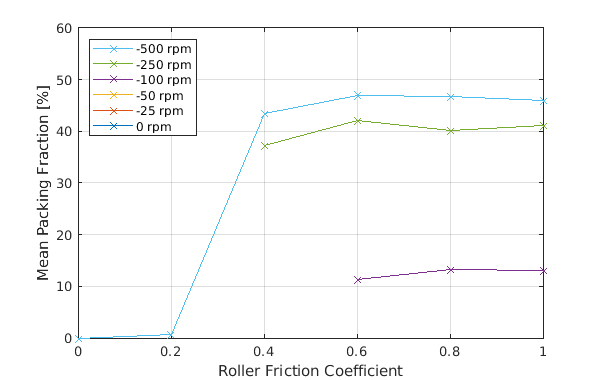}
         \caption{Spreading with counter-rotation}
         \label{fig:Rotation_low_sadh}
     \end{subfigure}
     \hfill
     \begin{subfigure}[b]{0.49\textwidth}
         \centering
         \includegraphics[trim = {8mm 0mm 10mm 2mm}, clip, scale=1, keepaspectratio=true, width=\textwidth]{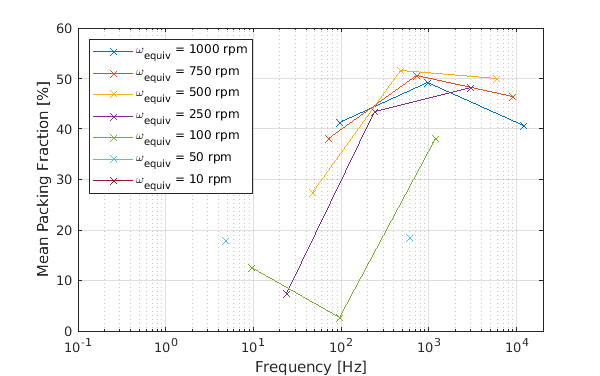}
         \caption{Spreading with angular oscillation}
         \label{fig:Oscillation_vmax_low_sadh}
     \end{subfigure}
     \caption{Mean packing fraction of powder layers spread over a substrate with a traverse velocity of v = $25 \frac{mm}{s}$ and friction coefficient of $\mu = 0.8$, where the substrate adhesion is reduced by a factor of 10; Note: data points where no powder was deposited not depicted, e.g., spreading with counter-rotational velocity of 0, -25, and -50~rpm yielded no powder layer}
     \label{fig:reduced sadh}
\end{figure}

Figure~\ref{fig:Rotation_low_sadh} shows the results of the equivalent study from Section~\ref{subsec:rotational spreading}, here with reduced substrate adhesion. The main insight is that for $\omega$ = -500 rpm and a roller friction coefficient of 0.4 or larger, the mean packing fraction is comparable to the case with the normal surface adhesion levels of the substrate. Specifically, for the case of $\omega$ = -500 rpm and $\mu$ = 0.6, the packing fraction is almost identical (47.0\% for $\gamma_{substrate, modified}$ vs 47.7\% for the default substrate surface adhesion). For friction coefficients $\mu < 0.4$ the powder is not spreadable anymore, a result that is independent of the rotational velocity.
The influence of roller rotation and friction coefficient seems to be significantly more important for a lower substrate adhesion value. For $\omega$ = -250 rpm, the achievable mean packing fraction is comparable to the $\omega$ = -100 rpm case in Section~\ref{subsec:rotational spreading}, however powder is not being spread below $\mu$ = 0.4. For rotational velocities below $\omega$ = -250 rpm, the layer packing fraction is either very low (around 10-15\% for $\omega$ = -100 rpm and $\mu \geq$ 0.6) or no powder is spread (all cases below $\omega$ = -100 rpm). In this case, the powder pile is simply pushed across the substrate without being deposited and no data point is displayed in Figure~\ref{fig:Rotation_low_sadh}.

For the spreading with angular oscillation, the results are similar. The friction coefficient is set to $\mu = 0.8$ for all angular oscillation experiments. For low comparable equivalent maximal rotational velocities, no powder is deposited (e.g., 10 rpm and one 50 rpm scenario which are therefore not plotted). The best spreading results on the other hand are comparable to the best spreading conditions in the unaltered simulations in Section~\ref{subsec:oscillation spreading}. For example, the best result is achieved with $\omega_{equivalent}$ = 500 rpm, amplitude A = 1$^{\circ}$, and frequency f = 4.77e-1 kHz, where the default case had a mean packing fraction of 52.2\% and the scenario with the reduced substrate adhesion achieves 51.5\%.

% In summary, while spreading becomes more challenging and in several cases infeasible, the spreading conditions that produce best results on a substrate that has similar cohesive properties as the powder also produce best results on a less cohesive substrate. Even as substrate conditions change significantly, the layer packing fraction stays very consistent which underlines the robustness of these spreading strategies and parameter choices.

\subsection{Spreading with a roller without surface adhesion}
\label{subsec:roller_no adhesion}

Besides influencing the surface roughness by changing the surface profile of the spreading implement -- in this work a roller -- practitioners could also change the material to affect the adhesion of particles to the tool. For blades, we have previously shown that a lower level of adhesion between particles and the tool is beneficial for spreading~\cite{Meier2019CriticalManufacturing}. Here, we look at the limiting case of $\gamma_{roller}=0$, meaning that particles do not adhere to the roller at all. While this case is unrealistic in reality, it is the limiting case showing what theoretical maximum improvement could be reached due to the choice of material. 
Figure~\ref{fig:roller_combo_plot} includes the results of two exemplary simulations where the adhesion for the roller $\gamma_{roller}$ is set to zero. The mean packing fraction is on the abscissa, with the ordinate showing the standard deviation of the packing fraction field. 
First, the best case of the pure counter-rotating simulation with v~=~$25 \frac{mm}{s}$ from Section~\ref{subsec:rotational spreading}, i.e., $\omega$~=~500 rpm and $\mu$~=~0.8 (blue cross), is rerun without roller adhesion (red cross). Second, the best case of the angular oscillation simulation with v~=~$25 \frac{mm}{s}$ from Section~\ref{subsec:oscillation spreading}, i.e., $\omega_{equivalent}$~=~500 rpm, amplitude A~=~1$^{\circ}$ and $\mu$~=~0.8 (blue circle), is rerun without adhesion to the roller (red circle).

Both simulations show no significant impact of the roller adhesion on the resulting powder layer quality. In the simple counter-rotating case, the reduced adhesion even slightly reduces the mean packing fraction from 48.1\% to 46.5\% while increasing uniformity in terms of a slightly lower standard deviation which decreases from 4.6\% to 3.9\%. For the oscillation simulations, the mean packing fraction also slightly decreases from 52.2\% to 50.7\% while the standard deviation of the packing fraction field stays almost the same with a minimal increase from 2.5\% to 2.6\%.

Summarizing these results, reducing the surface adhesion of the roller has no significant impact on the quality of the powder layer when using optimal parameters.

\subsection{Spreading with a flexible roller}
\label{subsec:flexible roller}

\begin{figure}
% \centering
%      \begin{subfigure}[b]{0.49\textwidth}
%          \centering
%          \includegraphics[trim = {8mm 0mm 10mm 2mm}, clip, scale=1, keepaspectratio=true, width=\textwidth]{Figs/Roller_combo_plot_v2.png}
     % \end{subfigure}
     % \hfill
     % \begin{subfigure}[b]{0.49\textwidth}
     %     \centering
     %     \includegraphics[trim = {8mm 0mm 10mm 2mm}, clip, scale=1, keepaspectratio=true, width=\textwidth]{Figs/Roller_combo_plot_v3.png}
     % \end{subfigure}
% \vspace{-12pt}
 \begin{center}
   \includegraphics[trim = {8mm 0mm 10mm 2mm}, clip, scale=1, keepaspectratio=true, width=0.5\textwidth]{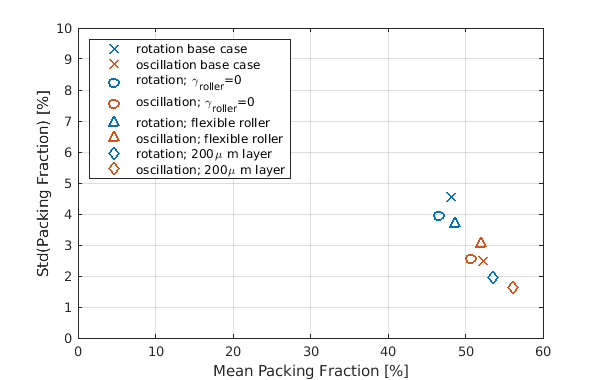}
 \end{center}
 % \vspace{-16pt}
 \caption{Mean and standard deviation of the packing fraction field for a variety of roller simulations with v~=~$25~\frac{mm}{s}$, $\omega$~=~-500~rpm and $\mu$~=~0.8 for simple counter-rotation simulations and v~=~$25~\frac{mm}{s}$, $\omega_{equivalent}$~=~500~rpm, amplitude A~=~1$^{\circ}$ and $\mu$~=~0.8 for angular oscillation simulations}
   % \vspace{-12pt}
 \label{fig:roller_combo_plot}
\end{figure}

Flexible blades have shown to be advantageous to mitigate powder layer defects such as streaking~\cite{PENNY2022Blade} due to aggregated particles or spatter, prevent tool damage in case of collisions with part edges, and reduce the peak forces the spreading tool exerts on the powder layer. These advantages are likely to be transferable to rollers made from or covered with soft material as well. For example, Tang et al.~\cite{TANG2022111489}, employ a flexible roller in an experimental study investigating co-rotational spreading with a recoating pressure of up to 2~MPa. The tool has a stainless steel core with a diameter of 5~mm and is coated with fluorine rubber (E-modulus = 7.84 MPa) with a thickness of 1.5~mm (outer diameter of 8~mm), aiming to achieve condensed layers of a powder with weak flowability (due to the low sphericity of the particles).

\begin{figure}
    \centering
    \includegraphics[width=\textwidth, trim={0 8cm 0 0}, clip, scale=1, keepaspectratio=true]{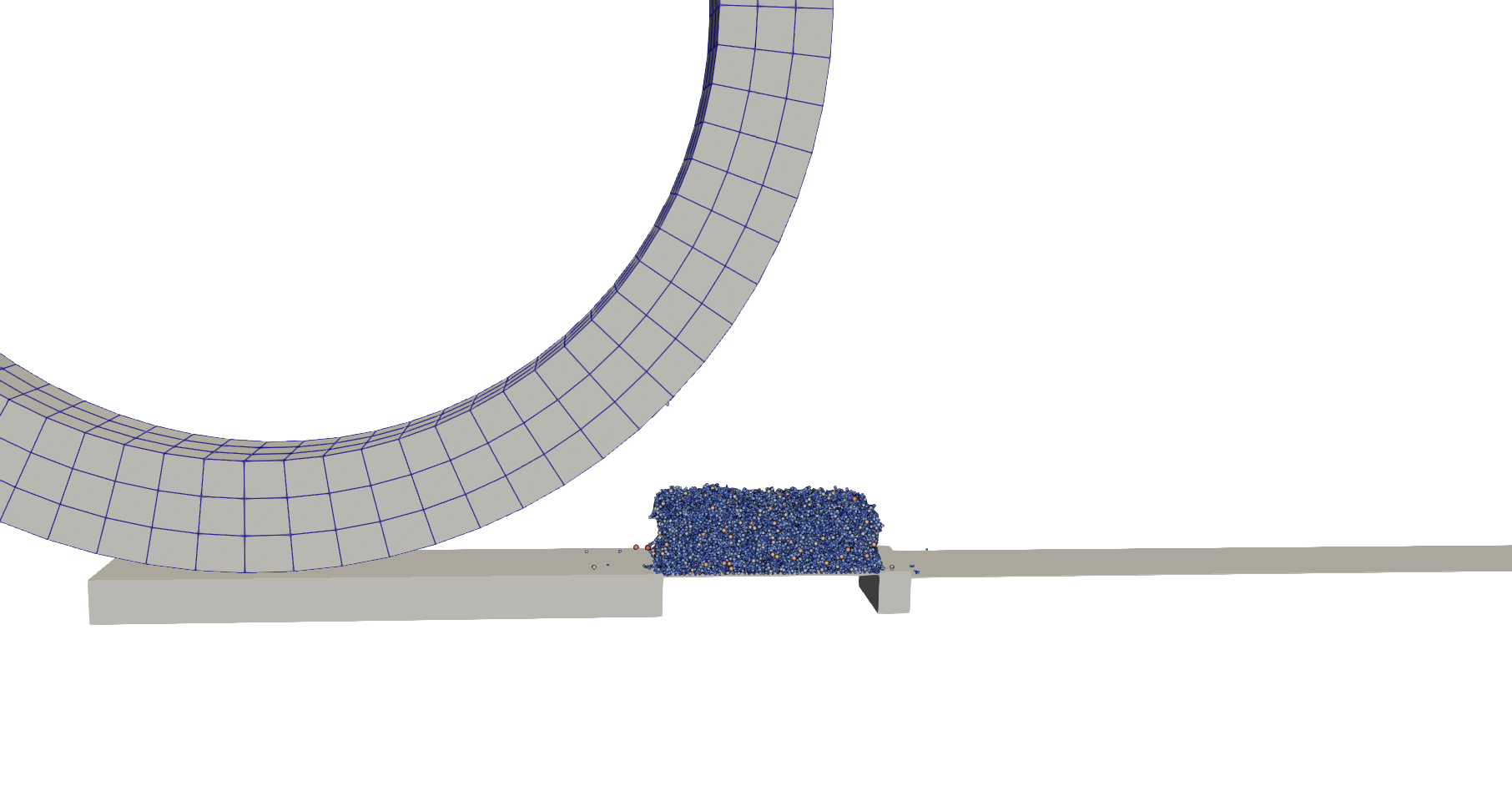}
    \caption{Simulation setup with the flexible roller where the rigid core is represented by the displacement-controlled inner surface whereas the 1~mm layer of compliant material is discretized by three finite element layers}
    \label{fig:flexible_roller}
\end{figure}

With the described integrated DEM-FEM framework, we simulate powder spreading employing a flexible, rubber-coated roller with a rigid core (see Figure~\ref{fig:flexible_roller}). Keeping the outer diameter of the roller at 10~mm, the roller has a rigid core of 8 mm diameter and is covered by a 1 mm layer of compliant material. In the simulations, the motion of the inner rigid core is controlled via displacement boundary conditions as employed throughout the previous sections. The outer (rubber) layer is modeled as a hyperelastic Saint Venant Kirchhoff material with an elastic modulus of 5.53~MPa and discretized with approximately~700 finite elements (FEM). In the radial direction, the elastic layer is represented by three finite element layers. This material property emulates a nitrile (Buna-N) O-ring stock (McMaster-Carr, 9700K16), that was previously used in \cite{PENNY2022Blade} for a flexible blade.

Exemplary spreading scenarios using the counter-rotational as well as the angular oscillation approach, both for v = $25 \frac{mm}{s}$ (see Section~\ref{subsec:rotational spreading}), are analyzed. For the counter-rotating scenario, the layer quality increases slightly when using the flexible roller, indicated by the similar packing fraction (48.6\% vs. 48.1\%) and the standard deviation of the packing fraction field decreases slightly (3.7\% vs. 4.6\%), compared to the base scenario from Section~\ref{subsec:rotational spreading} (blue triangle vs. blue cross in Figure~\ref{fig:roller_combo_plot}). Also for the angular oscillation approach, the influence of the coating is rather small. While the mean packing fraction is comparable as well (51.9\% vs. 52.2\%), the standard deviation of the packing fraction field increases slightly (3.0\% vs. 2.5\%) compared to the base scenario from Section~\ref{subsec:oscillation spreading} (red triangle vs. red cross in Figure~\ref{fig:roller_combo_plot}).
The fact that the layer quality between compliant and rigid roller is comparable is a good result, since the objective with using compliant tools is to utilize their significant practical advantages, e.g., in preventing tool damage in case of collisions or in mitigating singular effects such as streaking of very large particles or agglomerates~\cite{PENNY2022Blade}, which have not been considered in the present powder spreading simulations.

\begin{figure}
    \centering
    \includegraphics[width=0.6\textwidth, trim={0 0 0 0}, clip, scale=1, keepaspectratio=true]{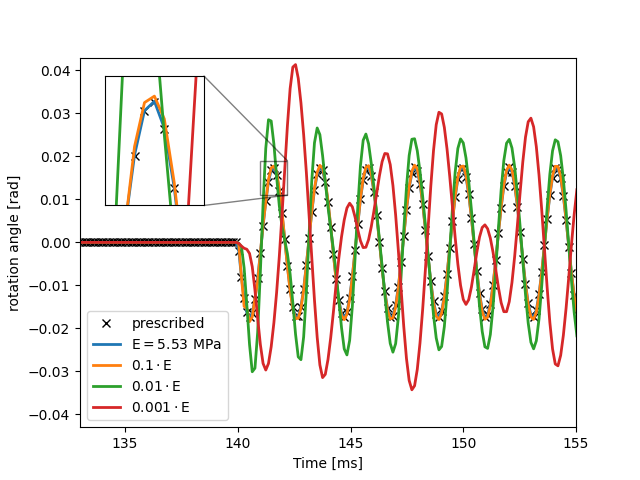}
    \caption{The angular displacement of a point on the roller surface over time for the oscillating roller simulations between 133 and 155~ms with different Young's moduli. As reference the prescribed rotation angle at the inner surface, i.e. of the rigid core, is visualized.}
    \label{fig:flexible_roller_rotation}
\end{figure}

To study the shear deformation of the rubber coating during the rotation, Figure~\ref{fig:flexible_roller_rotation} shows the rotation angle $\alpha$ (see Eq.~\eqref{eq:oscillation_alpha}) of one point on the outer surface of the roller over time. Using the Young's modulus ($E=5.53$~MPa) of nitrile (Buna-N) O-ring stock (blue curve in Fig.~\ref{fig:flexible_roller_rotation}), the rubber coating follows the prescribed rotation almost exactly. Even a reduction of the Young's modulus by a factor of 10 (orange curve) shows only a negligible difference to the prescribed rotation. Only when reducing the Young's modulus by an unrealistic factor of 100 the outer surface overshoots the prescribed rotation and, eventually, at a reduction of 1000 the rubber coating is so soft that it cannot follow the prescribed rotation anymore. Thus, a rubber-coated roller is expected to combine both the stiffness to appropriately follow the applied kinematics while being compliant enough to have the benefits of spreading with compliant blades. To confirm this hypothesis, future work could assess the effect of perturbations on the roller surface to see if the benefits hold true given the difference in structural stiffness between a fully compliant blade and a coated steel roller.

\subsection{Influence of layer thickness}

In this section, exemplary spreading scenarios using the counter-rotational as well as the angular oscillation approach are analyzed for an increased layer thickness of 200~$\mu m$ (unscaled with D90~=~43.4~$\mu m$ and cutoff diameter of 88~$\mu m$, corresponding to $\sim$80~$\mu m$ for the 0-20~$\mu m$ powder). If successfully fused, a thicker layer is an effective means to increase throughput when building up a part -- when resolution requirements permit the thicker layers.

As expected, the increased layer thickness significantly increases layer quality for both spreading approaches. Figure~\ref{fig:roller_combo_plot} shows the results of the two exemplary simulations in the diamond shape. Similar to the previous sections that reference Figure~\ref{fig:roller_combo_plot}, the blue markers indicate the pure counter-rotating approach (v~=~$25 \frac{mm}{s}$, $\omega$~=~500 rpm and $\mu$~=~0.8) while the red markers represent the angular oscillation simulations (v~=~$25 \frac{mm}{s}$, $\omega_{equivalent}$~=~500 rpm, amplitude A~=~1$^{\circ}$ and $\mu$~=~0.8). 

For the rotational case, the mean packing fraction increases from 48.1\% to 53.5\%, while the standard deviation decreases significantly from 4.6\% to 2.0\%. For the angular oscillation scenario, the mean packing fraction even increases from 52.2\% to 56.1\%. The standard deviation decreases even further from 2.5\% to 1.6\%.

This significant improvement in layer quality confirms previous studies~\cite{yao2021dynamic,nan2020a, Meier2019CriticalManufacturing, zhang2020, Mindt2016}. Two key factors contribute to this increased packing fraction. First, in a thicker layer, powder flow through the spreading gap is improved and the impact of singular effects such as locking or streaking of very large particles or agglomerates is decreased. Second, the impact of the flat bottom, which prohibits optimal particle arrangements at the boundary, is decreased. The higher packing fraction and increased uniformity is beneficial for subsequent melting or binder absorption, rendering thick layers of cohesive powders a potentially promising avenue to increasing throughput in metal AM.

%The strong decrease in standard deviation can be traced back to the fact that in a thicker layer, particles have more options to arrange in a way that is beneficial to a smooth surface. Once again, large particles that can create a local non-uniformity in a thin powder layer have less of an impact in a layer that is thick compared to their diameter.

\subsection{Impact of roller diameter}

\begin{figure}[h!]
     \centering
     \begin{subfigure}[b]{0.49\textwidth}
         \centering
         \includegraphics[trim = {8mm 0mm 10mm 2mm}, clip, scale=1, keepaspectratio=true, width=\textwidth]{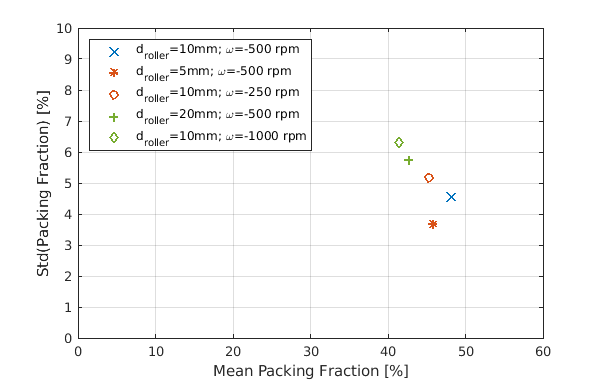}
         \caption{Spreading with a counter-rotating roller with traverse velocity of v = $25 \frac{mm}{s}$}
         \label{fig:roller size rotation}
     \end{subfigure}
     \hfill
     \begin{subfigure}[b]{0.49\textwidth}
         \centering
         \includegraphics[trim = {8mm 0mm 10mm 2mm}, clip, scale=1, keepaspectratio=true, width=\textwidth]{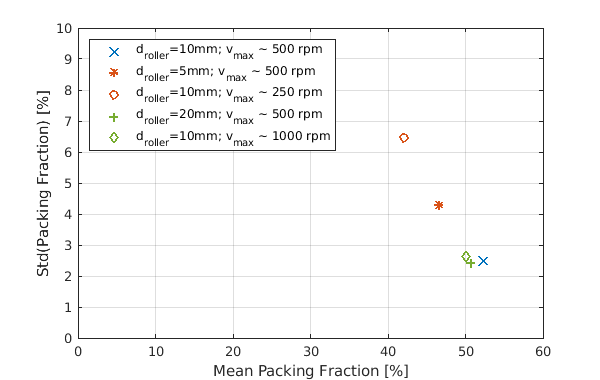}
         \caption{Spreading with angular oscillation and a traverse velocity of v = $25 \frac{mm}{s}$}
         \label{fig:roller size oscillation}
     \end{subfigure}
     \caption{Comparison of simulations with different roller diameter (base case: d$_{roller}=10$~mm), equivalent angular velocities, and comparable roller surface velocities}
     \label{fig:roller size}
\end{figure}

For a practitioner using a roller spreading implement, the roller diameter is another important parameter that can be optimized. To investigate this aspect, we look at the impact of changing the roller diameter, specifically comparing the diameters 5~mm and 20~mm to the base case of 10~mm. 
Again, in a similar fashion as in the previous studies, a base case for the rotational (v~=~$25 \frac{mm}{s}$, $\omega$~=~-500~rpm and $\mu$~=~0.8) and the angular oscillation (v~=~$25 \frac{mm}{s}$, $\omega_{equivalent}$~=~500~rpm, amplitude A~=~1$^{\circ}$ and $\mu$~=~0.8) spreading is compared to corresponding simulations with varied roller diameter.
For the 20~mm roller simulations, due to the larger size of the roller, the number of particles is increased to 105,000. This allows for a heap of powder in front of the roller that resembles actual spreading more realistically. 
Given the otherwise unchanged simulation parameters, the surface velocity of the larger roller is significantly higher (and for the smaller roller significantly smaller) compared to the d=10~mm roller ($\omega$~=~-500~rpm). The velocity component originating from the angular motion doubles/halves given the same angular velocity. E.g., for the 20~mm roller the angular motion of -500~rpm is comparable to an angular velocity of -1000~rpm of the roller with a diameter of 10~mm; vice versa for the 5~mm roller to an angular velocity of -250~rpm. Thus, for better comparison, these cases are also included in Figure~\ref{fig:roller size}. 

For both the rotational simulations (see Figure~\ref{fig:roller size rotation}) as well as the angular oscillation approach (see Figure~\ref{fig:roller size oscillation}), the base scenario of d=10~mm and -500~rpm achieves the highest packing fraction.

The powder layer density decreases with the larger roller as well as the smaller roller, compared to the $\omega$~=~-500~rpm scenario (see Table~\ref{table:large roller results}). E.g., for the pure rotational simulation, the increased roller diameter leads to a decrease of the packing fraction from 48.1\% to 42.7\%, while the standard deviation increases from 4.6\% to 5.8\%. For the angular oscillation, the packing fraction slightly decreases from 52.2\% to 50.6\% while the standard deviation remains almost unchanged at 2.4\% (vs. 2.5\% in the base case). This observation can be explained by the fact, that the 500~rpm case of the d=10~mm roller has been identified as the optimal angular velocity, resulting also in optimal surface velocities, in the previous sections. In contrast, the rollers with the smaller / larger diameter lead to surface velocities that are smaller / larger than the optimal surface velocity of the reference case with $\omega$~=~-500~rpm for d=10~mm. 

However, when comparing the different roller diameter simulations with an equivalent circumferential velocity instead of the similar angular velocity, the simulations with similar circumferential velocity generally result in very similar powder layers. E.g., comparing the simulation with d=20~mm and $v_{max}\sim$500~rpm and the simulation with d=10~mm and $v_{max}\sim$1000~rpm for the angular oscillation, the packing fraction and standard deviation are almost identical (mean of 50.6\% for d=20~mm vs. 50.1\% for d=10~mm and standard deviation of 2.4\% vs. 2.6\% respectively, see Table~\ref{table:large roller results}.

% \begin{table}
% \centering
% \caption{Mean packing fraction $\bar{\Phi}$ and standard deviation $std(\Phi)$ for simulations comparing different roller sizes, all with v~=~$25 \frac{mm}{s}$}
% \label{table:large roller results}
% \begin{tabular}{ |l|c|r|r|r| } 
%  \hline \raggedleft
%  & &d=10~mm, $\omega$~=~-500~rpm &d=10~mm, $\omega$~=~-1000~rpm & d=20~mm, $\omega$~=~-500~rpm\\ 
%  \hline
%  \multirow{2}*{Rotation}&$\bar{\Phi}$   &48.1\% &41.4\% &42.7\% \\ 
%  \cline{2-5}    &$std(\Phi)$   &4.6\% &6.3\% &5.8\% \\ 
%  \hline
%  \multirow{2}*{Oscillation}&$\bar{\Phi}$   &52.2\% &50.1\% &50.6\% \\ 
%  \cline{2-5}    &$std(\Phi)$   &2.5\% &2.6\% &2.4\% \\  
%  \hline
% \end{tabular}
% \end{table}

\begin{table}
\centering
\caption{Mean packing fraction $\bar{\Phi}$ and standard deviation $std(\Phi)$ for simulations comparing different roller sizes, all with v~=~$25 \frac{mm}{s}$}
\label{table:large roller results}
\begin{tabular}{ |l|c|c|c|c| } 
 \hline \raggedleft
 & \multicolumn{2}{c|}{Rotation} &\multicolumn{2}{c|}{Oscillation}\\ 
 \cline{2-5}
 &$\bar{\Phi}$   &$std(\Phi)$ &$\bar{\Phi}$   &$std(\Phi)$ \\ 
 \hline
 d=10~mm, $\omega$~=~-500~rpm &48.1\%   &4.6\% &52.2\% &2.5\% \\ 
 \hline
   d=5~mm, $\omega$~=~-500~rpm &45.8\%   &3.7\% &46.6\% &4.3\% \\  
 \hline
 d=10~mm, $\omega$~=~-250~rpm &45.2\%   &5.2\% &42.0\% &6.5\% \\ 
 \hline
 d=20~mm, $\omega$~=~-500~rpm &42.7\%   &5.8\% &50.6\% &2.4\% \\ 
 \hline
 d=10~mm, $\omega$~=~-1000~rpm &41.4\%   &6.3\% &50.1\% &2.6\% \\  
 \hline
\end{tabular}
\end{table}

Given that the powder layer quality is comparable for the equivalently chosen circumferential velocity, we conclude that the changed geometry has relatively little impact compared to the kinematic spreading parameters.

\section{Conclusion}

Using a coupled DEM-FEM computational model, this work investigated the spreading of very cohesive powders, as required e.g. in LPBF or BJ additive manufacturing, by means of a roller implement. Compared to spreading with a blade, using a counter-rotating roller can create thin powder layers of high quality with fine, cohesive powders when optimizing the kinematic parameters. This indicates that there is a material-specific process window for the deposition of these cohesive powders. For the investigated highly cohesive 0-20~$\mu m$ Ti-6Al-4V powder, counter-rotating angular velocities of 500~rpm, a sufficiently high friction coefficient of the roller surface of $\mu \geq 0.4$ and moderate traverse velocities of around 10-25~$\frac{mm}{s}$ achieved best results in terms of a high packing fraction with low variance. For angular velocities $<$500~rpm (and also for low $\mu$) insufficient kinetic energy is induced into the powder as to effectively break cohesive bonds, while for angular velocities $>$500~rpm the induced kinetic energy is too high, leading to significant particle ejection and a pronounced "dusty zone" in front of the roller, both resulting in reduced packing fraction as compared to the best case with $\omega = 500$~rpm. Based on these results we propose a process window for spreading of cohesive powders, that defines a zone of layers with a high packing fraction within the parameter space of traverse velocity and angular velocity.

As experimentally shown by Penny et al.~\cite{PENNY2022Roller}, a high manufacturing precision for the spreading apparatus is required in order to avoid a wave-like surface profile coming from the runout of the rotating roller. To mitigate the difficulty of manufacturing with such high precision, roller-based spreading with angular oscillation is proposed. Here, oscillation amplitudes in the order of the diameter of the largest particles in the powder size distribution (corresponding to an amplitude of approximately 1$^{\circ}$ when considering angular oscillations with the given roller dimensions and powder size) showed best results. Interestingly, angular oscillation yielded the best layer quality when the oscillation frequency was chosen such that the resulting maximal circumferential velocity on the roller surface is comparable to the counter-rotating case with optimal angular velocity $\omega = 500$~rpm. Critically, both of these spreading approaches, i.e., counter-rotation as well as angular oscillation, are shown to be very robust with respect to varying substrate conditions (e.g., reduced substrate adhesion by one order of magnitude), which are likely to occur in LBPF or BJ, where substrate characteristics are the result of a complex multi-physics (i.e., powder melting or binder infiltration) process. Specifically, angular oscillation kinematics showed a high level of robustness regarding kinematic parameters (i.e., high layer quality across a wide range of frequencies) as well as substrate conditions.

While the present study considered the most challenging case of rather thin layers (layer thickness of approximately the size of the largest particle) of highly cohesive powers, it was demonstrated that higher packing fraction levels can be achieved for thicker layers, an effect that confirms previous studies. On the other hand, parameters such as the roller diameter (for a fixed value of the maximal circumferential velocity on the roller surface) and the powder-to-roller adhesion (controllable, e.g., through the roughness and material of the roller surface) were shown to have less influence on the layer quality as compared to the roller kinematics, at least in the regime of optimal angular / oscillation velocities. 

% In conclusion, the proposed angular oscillation patterns on roller-like geometries with oscillation amplitudes in the order of the largest particles in the powder size distribution, a sufficiently rough surface of the spreading tool (coefficient of friction above 0.4) and a low to moderate traverse velocity are recommended. Optimal oscillation frequencies have to be determined based on the cohesiveness of the given powder material such that the induced kinetic energy is high enough to break adhesive bonds but not too high, which would result in excessive particle ejection. In sum, the following key benefits can be stated:
% \begin{itemize}
%     \item high and uniform packing densities beyond 50 percent even for very thin layers of highly cohesive powders
%     \item robustness w.r.t. varying substrate conditions (e.g., reduced substrate adhesion)
%     \item robustness w.r.t. deviations from the optimal oscillation frequency (more robust than pure counter-rotation)
%     \item no runout problems as compared to pure rotation kinematics
%     \item oscillation can be extended to non-roller geometries
% \end{itemize}

Apart from different roller kinematics, also first studies on the usage of composite recoating tools, e.g. rubber-coated rollers, have been performed. Fully compliant recoating tools (e.g. made of rubber) are well-known to reduce the risk of tool damage and quality-degrading singular effects such as streaking of large particles, but typically the achievable manufacturing tolerances are not sufficient to allow, e.g., for counter-rotation roller spreading. A promising alternative has been proposed in this work by applying oscillatory kinematics (which are less restrictive with respect to manufacturing tolerances) to a rubber-coated steel roller combining the advantages of stiff and compliant spreading tools (i.e., kinematic precision and robustness). Our preliminary studies confirmed that such rubber-coated rollers allow to precisely control oscillatory surface kinematics and to achieve a comparable packing density and layer uniformity as with steel rollers as long as no singular effects such as particle streaking occur (which rarely happens on the domain sizes typically considered in computational modelling). Therefore, the usage of such rubber-coated rollers, and in particular the analysis of their robustness with respect to collisions and streaking, are considered as very promising avenues of future research. In addition, our future research will also focus on an experimental validation of the proposed oscillatory spreading approaches.

\section{Acknowledgements}

Financial support at MIT was provided by a MathWorks MIT Mechanical Engineering Fellowship (to R.W.) as well as by the National Science Foundation (Award EEC-1720701, subcontracted from the University of Illinois at Urbana-Champaign). P.P., C.M., and W.W. acknowledge funding by the Deutsche Forschungsgemeinschaft (DFG, German Research Foundation) within project 414180263. C.M. acknowledges funding by the European Union (ERC starting grant "ExcelAM"; project number: 101117579). We also thank Ryan W. Penny (MIT) for providing valuable insights from his experimental work with fine metal powders.

\newpage
%\section*{References}
\bibliographystyle{elsarticle-num}
\typeout{}
\bibliography{refs.bib}
\newpage

\beginsupplement
\FloatBarrier
\section{Analysis of variation between simulations with the same parameters}

\begin{figure}[h!]
     \centering
     % \begin{subfigure}[b]{0.49\textwidth}
     %     \centering
     %     \includegraphics[trim = {8mm 0mm 10mm 2mm}, clip, scale=1, keepaspectratio=true, width=\textwidth]{Figs/Roller_oscillation_amp_study_v2.png}
     %     \caption{Repeat of Figure~\ref{fig:Oscillation_amp} for comparison: Results plotted by oscillation amplitude}
     %     \label{fig:Oscillation_amp_supp}
     % \end{subfigure}
     % \hfill
     \begin{subfigure}[b]{0.49\textwidth}
         \centering
         \includegraphics[trim = {8mm 0mm 10mm 2mm}, clip, scale=1, keepaspectratio=true, width=\textwidth]{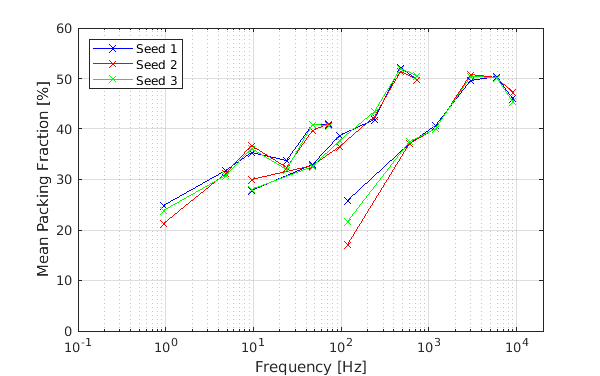}
         % \caption{Results of Figure~\ref{fig:Oscillation_amp} for different seeds}
         \label{fig:seed_comparison}
     \end{subfigure}
     \caption{Results of Figure~\ref{fig:Oscillation_amp} for different random powder samples}
     \label{fig:seed_vs_original}
\end{figure}

% \section{Computational Modeling Parameters}

% \begin{table}[htb]
% \centering
% \renewcommand{\arraystretch}{0.7}
% \begin{tabular}{ l l l } 
%  \hline
%  Parameter & Value & Unit \\ 
%  \hline
%  \multirow{2}{*}{Density}       &Ti-6Al-4V: 4430      &kg/m$^3$     \\ 
%       &Al-10Si-Mg: 2670      &kg/m$^3$   \\ 
%  Penalty parameter      &0.13      &N/m   \\ 
%  Poisson's ratio     &0.342         &- \\
%  Coefficient of friction     &0.4         &- \\
%  Coefficient of rolling friction     &0.07         &- \\
%  Coefficient of restitution     &0.4         &- \\
%  Surface energy     &varied: 0.02-2.56         &mJ/m$^2$ \\
%  \hline
%  Log-normal particle size distribution:     &         & \\
%  Median     &13.4968         &$\mu$m \\
%  Sigma     &0.2253         &- \\
%  Minimum cutoff radius     &10.1117         &$\mu$m \\
%  Maximum cutoff radius     &44         &$\mu$m \\
%  \hline
%  \end{tabular}
% \caption{DEM model parameters}
% \label{tab:DEMtarameters}
% \end{table}

\newpage
\section{Full parameter study for counter-rotational spreading}

\begin{figure}[h!]
     \centering
     \begin{subfigure}[b]{0.49\textwidth}
         \centering
         \includegraphics[trim = {8mm 0mm 10mm 2mm}, clip, scale=1, keepaspectratio=true, width=\textwidth]{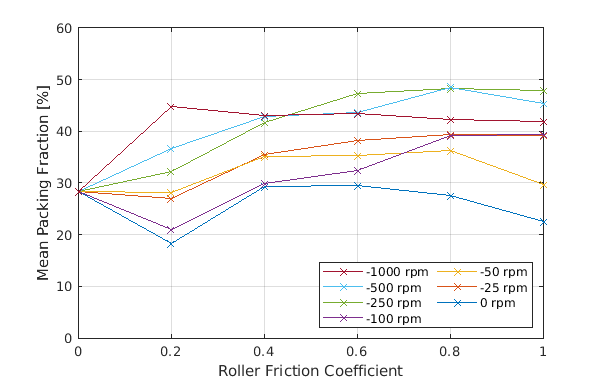}
         \caption{Traverse velocity v = $5 \frac{mm}{s}$}
         \label{fig:Rotational study 5mms}
     \end{subfigure}
     \hfill
     \begin{subfigure}[b]{0.49\textwidth}
         \centering
         \includegraphics[trim = {8mm 0mm 10mm 2mm}, clip, scale=1, keepaspectratio=true, width=\textwidth]{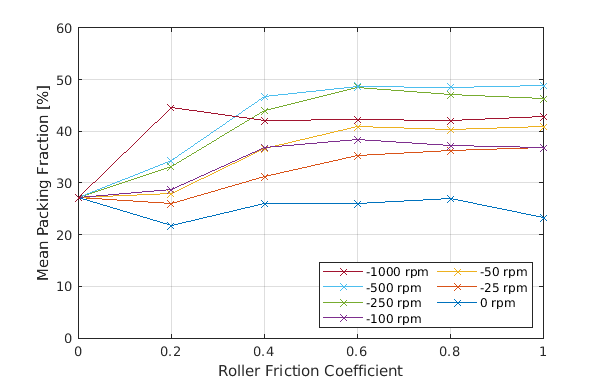}
         \caption{Traverse velocity v = $10 \frac{mm}{s}$}
         \label{fig:Rotational study 10mms}
     \end{subfigure}
     \vfill
     \begin{subfigure}[b]{0.49\textwidth}
         \centering
         \includegraphics[trim = {8mm 0mm 10mm 2mm}, clip, scale=1, keepaspectratio=true, width=\textwidth]{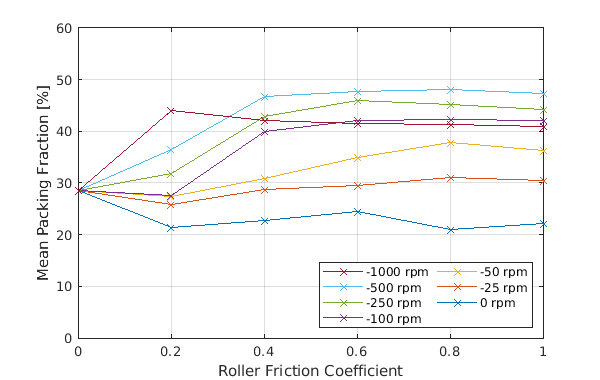}
         \caption{Traverse velocity v = $25 \frac{mm}{s}$}
         \label{fig:Rotational study 25mms}
     \end{subfigure}
     \hfill
     \begin{subfigure}[b]{0.49\textwidth}
         \centering
         \includegraphics[trim = {8mm 0mm 10mm 2mm}, clip, scale=1, keepaspectratio=true, width=\textwidth]{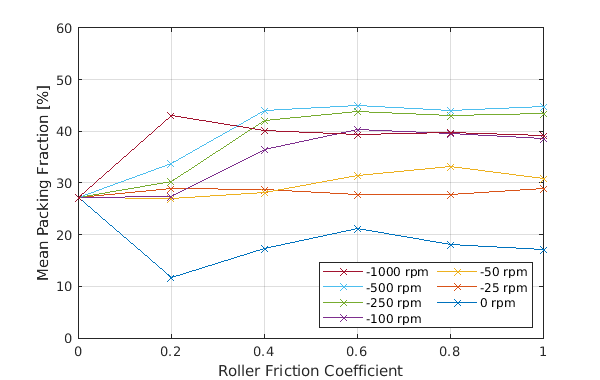}
         \caption{Traverse velocity v = $50 \frac{mm}{s}$}
         \label{fig:Rotational study 50mms}
     \end{subfigure}
     \caption{Mean packing fraction of powder layers spread with a counter-rotating roller for varying rotational velocity, roller friction coefficient and traverse velocity}
     \label{fig:rotational study supp}
\end{figure}

\end{document}